\documentclass[english,english]{article}
\usepackage{times}
\usepackage[T1]{fontenc}
\usepackage[latin1]{inputenc}
\usepackage{a4wide}
\usepackage{amsmath}
\usepackage{amssymb}

\makeatletter
\allowdisplaybreaks[1]
\newcommand{\myMathBox}[1]{%
\vspace{\abovedisplayskip}%
\noindent%
\fbox{\parbox{1.1\linewidth}{#1}}%
\vspace{\belowdisplayskip}%
}

\usepackage{babel}
\makeatother
\begin{document}

\title{Deriving the sampling errors of correlograms for general white noise}

\author{T. D. Carozzi and A. M. Buckley\\
\emph{Space Science Centre, Sussex University, Brighton BN1 9QT, United
Kingdom}\texttt{}~\\
T.Carozzi@sussex.ac.uk}

\date{20 May 2005}

\maketitle
\begin{abstract}
We derive the second-order sampling properties of certain autocovariance
and autocorrelation estimators for sequences of independent and identically
distributed samples. Specifically, the estimators we consider are
the classic lag windowed correlogram, the correlogram with subtracted
sample mean, and the fixed-length summation correlogram. For each
correlogram we derive explicit formulas for the bias, covariance,
mean square error and consistency for generalised higher-order white
noise sequences. In particular, this class of sequences may have non-zero
means, be complexed valued and also includes non-analytical noise
signals. We find that these commonly used correlograms exhibit lag
dependent covariance despite the fact that these processes are white
and hence by definition do not depend on lag.
\end{abstract}

\section{Introduction}

Serial correlation, although not as popular as its Wiener-Khintchine
theorem equivalent, spectral analysis, is still of fundamental importance
in the analysis of time series, and is used in a wide range of applications
ranging from radio astronomy \cite{Smolders99} and radar scatter
from random media \cite{Schiffler96} to DNA sequencing \cite{Bertorelle95}
and wave-particle interaction instruments for space plasma research
\cite{Gough03}.

One of the most fundamental problems in time series analysis is to
discern if the samples in a given series are independent. Under stationary
conditions, this problem is most naturally dealt with using the autocorrelation
sequence (ACS) of the series since it measures the correlation between
samples in the series which are separated by some interval of time
known as the lag. In practice, we can only estimate an ACS based on
a finite number of data samples. So if we are to successfully detect
the independence of the samples we must know the sampling errors of
the ACS estimator for \emph{independent and identically distributed}
(IID) sample sequences, which are by definition independent. For our
purposes of deriving the sampling errors it turns out that we can
in fact broaden slightly the class of sequences from IID sequences
to higher-order white noise or, as we will sometimes call it, general
white noise.

It is of course the sampling errors or sampling properties that limit
how well we can determine sample independence from the estimated ACS
and it is therefore an important issue to explicitly provide them
for general white noise sequences. General sampling properties of
ACSs were first studied in \cite{Bartlett46}. Despite considerable
subsequent attention, it seems the important special case of general
IID sample sequences has not been fully exhausted. For instance, the
sampling properties of autocovariance estimates of \emph{complex}
valued noise has not been published and neither has the case of autocorrelation
estimates of \emph{nonzero mean} noise series. To be clear, the difference
between what we call autocorrelation and autocovariance is that in
autocovariance the mean is explicitly subtracted from data samples
while in autocorrelation it is not. Our use of these terms conforms
with \cite{Bendat00}. We will use the term \emph{correlogram} as
a collective term for estimators of autocovariance or autocorrelation
functions when there is no need to distinguish them.

In this paper we derive the sampling properties of correlograms for
complex valued higher-order white noise with arbitrary mean. The specific
correlograms we consider are the well known lag windowed autocorrelation
estimator, and the lag windowed autocovariance estimator in which
the sample mean is subtracted from the data. To contrast with the
ubiquitous lag windowed correlograms, we also consider the fixed-length
summation autocorrelation estimator.

The sampling properties of the estimators presented here are derived
to second-order. This includes the bias, the covariance and the mean-square-error
(MSE) and also the variance and consistency.

\section{Definitions and conventions}

In this section we introduce some basic definitions and notational
conventions.

In practice most correlograms are based on sequences of data samples
$Z[n]$ where $n\in\mathbb{Z}=\{0,\pm1,\pm2,...\}$. Here we denote
the sequence by putting a square bracket around the independent variable
to distinguish them from functions such as $Z(t)$ $t\in\mathbb{R}$
where the independent variable is continuous. The correlogram of sequences
will also be a sequence so we favor the terms autocorrelation sequence
(ACS), denoted $R[l]$ and autocovariance sequence (ACVS), denoted
$C[l]$, over the often used terms autocorrelation function (ACF),
denoted $R(l)$ and autocovariance function (ACVF), denoted $C(l)$.

Since we will be dealing with statistical estimators we will need
to distinguish between ensemble, or population, quantities and the
estimates, or samples, of the quantities. We will put a hat accent
on the symbol used for the estimators of a quantity so, for instance,
$\hat{R}$ is an estimate of $R$.

The distinction between the ACS and the ACVS is as follows. We define
the ACS as\[
R_{ZZ}[l]:=\mathrm{E}\left\{ Z^{*}[k]Z[k+l]\right\} ,\quad l,k\in\mathbb{Z}\]
while the ACVS is defined\[
C_{ZZ}[l]:=\mathrm{Cov}\left\{ Z[k],Z[k+l]\right\} :=\mathrm{E}\left\{ Z^{*}[k]Z[k+l]\right\} -\mathrm{E}\left\{ Z^{*}[k]\right\} \mathrm{E}\left\{ Z[k+l]\right\} \]
 where $E\{\cdot\}$ is the expectation operator, $\mathrm{Cov}\left\{ \cdot,\cdot\right\} $
is the covariance operator, $l$ is the lag, and $Z^{*}[k]$ is the
complex conjugate of $Z[k]$. Thus the difference between the ACS
and ACVS is that, in the case of the ACVS, the mean of the sequence
is explicitly subtracted out, or removed, while it is not removed
in ACS. Note that $C_{\cdot\cdot}$, $R_{\cdot\cdot}$, and, $\mathrm{Cov}\left\{ \cdot,\cdot\right\} $,
are all defined so that their first argument becomes complex conjugated,
this so that the default product between two unconjugated arguments
will be a hermitian form. 

The definition of the ACS varies in the literature with regards to
the placement of the complex conjugation. Here, it is defined in such
a way that if $Z[k]$ was the complex harmonic sequence $\exp(ifk)$,
the phase of its corresponding ACS would increase with increasing
lag, viz. $\exp(-ifk)\exp(if(k+l))=\exp(ifl)$.

\section{Generalised white noise}

General white noise sequences, in the sense we will use here, are
a less restrictive version of IID distributed samples in which the
cross-moments of any two different samples in the sequence are zero
in the lowest orders but may be non-zero at a higher order. This is
different from an IID since all their cross-moments are zero. It is
also slightly more general that higher-order white noise since we
allow for a non-zero mean and we also allow for a broader structure
of complex moments. The reason for considering these process is driven
entirely by the sampling error formulas themselves: the generalised
white noise is the least restrictive class of sequences with a flat
ACVS.

\subsection{Defining conditions of fourth-order general white noise}

For the purposes of determining the second-order sampling properties
of second-order lagged-moment estimates we need to know the statistical
lagged-moments of the signal only to fourth-order. In what follows,
we introduce a general white noise sequence, which we denote $\epsilon[n]$,
with the following defining properties: the first-order lagged moment
is\begin{equation}
E\{\epsilon[n]\}=\mu\end{equation}
where $\mu$ is the mean.

The ACVS of $\epsilon[n]$ is zero for all non-zero lags,

\begin{equation}
C_{\epsilon\epsilon}[l]=\mathrm{Cov}\left\{ \epsilon^{'}[k],\epsilon^{'}[k+l]\right\} =\mathrm{Cov}\left\{ \epsilon^{'*}[k],\epsilon^{'*}[k+l]\right\} =\mathrm{E}\left\{ \epsilon^{'*}[k]\epsilon^{'}[k+l]\right\} =\sigma^{2}\delta_{0,l},\quad k,l\in\mathbb{Z}\label{eq:HermCovOfGenWhiteNoise}\end{equation}
where $\sigma^{2}$ is the (hermitian) variance and $\delta_{a,b}$
is the Kronecker delta function. $\delta_{a,b}$ is equal to one if
$a=b$ and zero otherwise, hence $\delta_{0,l}$ represents the zero
lag, $l=0$. Equation (\ref{eq:HermCovOfGenWhiteNoise}) expresses
an important property of white noise, that its autocovariance function
is zero for all nonzero lags. It is precisely this property which
can be exploited of to test if a sequence is independent or not. Unfortunately
as we shall see, all autocovariance estimates will in practice be
distributed around zero and this distribution is specified according
to the estimators sampling properties. 

The ordinary ACVS for complex sequences is a hermitian bilinear form.
Since we are dealing with possibly complex valued sequences, the full
description of the second-order properties requires also another lagged
bilinear form, namely the lagged quadratic form covariance \begin{equation}
\mathrm{Cov}\left\{ \epsilon^{'*}[k],\epsilon^{'}[k+l]\right\} =\mathrm{E}\left\{ \epsilon^{'}[k]\epsilon^{'}[k+l]\right\} =m_{2}\delta_{0,l}=s^{2}\exp(i\theta_{2})\delta_{0,l}\label{eq:QuadCovOfGenWhiteNoise}\end{equation}
where $m_{2}$ is what we will call the quadratic variance. We call
the left-hand side of (\ref{eq:QuadCovOfGenWhiteNoise}) the quadratic
ACVS in contrast to the hermitian ACVS of the left-hand side of (\ref{eq:HermCovOfGenWhiteNoise}).
An important property of the quadratic ACVS is that it depends on
whether the signal is analytic or not. It is identically zero for
random stationary analytical signals \cite{Bendat00}. The quadratic
ACVF has, to the best of the authors knowledge, not been explicitly
investigated in the literature. This is because it is usually argued
that only analytical signals are of practical importance. This is,
however, does not mean the quadratic ACVS is without interest. In
terms of the second-order sampling properties derived here, the quadratic
ACVF is precisely what distinguishes purely real (non-imaginary) signals
from complex signals. For purely real signals the quadratic ACVS is
not zero, it is equivalent to the hermitian ACVS.

The third-order lagged moment is\begin{equation}
E\left\{ \epsilon^{'*}[k]\epsilon^{'}[k+l]\epsilon^{'}[k']\right\} =\kappa_{3}\delta_{kk'}\delta_{0l}\end{equation}
where $\kappa_{3}$ is the third-order cumulant of the zero lag and
is related to the skewness of distribution. Finally, the fourth-order
lagged moments of the white noise is\begin{align}
 & \mathrm{Cov}\left\{ \epsilon^{'*}[k]\epsilon^{'}[k+l],\epsilon^{'*}[k']\epsilon^{'}[k'+l']\right\} \nonumber \\
= & (\mu_{4}-\sigma^{4})\delta_{kk'}\delta_{0l}\delta_{0l'}+\sigma^{4}\delta_{kk'}\delta_{ll'}(1-\delta_{0l}\delta_{0l'})+|m_{2}|^{2}\delta_{l,k'-k}\delta_{-l,l'}(1-\delta_{0l}\delta_{0l'})\nonumber \\
= & (\mu_{4}-2\sigma^{4}-|m_{2}|^{2})\delta_{kk'}\delta_{0l}\delta_{0l'}+\sigma^{4}\delta_{kk'}\delta_{ll'}+|m_{2}|^{2}\delta_{l,k'-k}\delta_{-l,l'}\nonumber \\
= & \kappa_{4}\delta_{kk'}\delta_{0l}\delta_{0l'}+\sigma^{4}\delta_{kk'}\delta_{ll'}+s^{4}\delta_{k',k+l}\delta_{-l,l'}\end{align}
where $\mu_{4}$is the fourth-order central moment and$\kappa_{4}=\mu_{4}-2\sigma^{4}-|m_{2}|^{2}$
is the fourth-order cumulant. Fourth-order moments are related to
the kurtosis of the distribution.

To summarise, we specify a noise sequence $\epsilon[\cdot]$ which
is given by a minimum of 5 low order cumulants, namely, the mean,
variance, complex variance, third- and fourth-order cumulants. This
is the least restrictive white noise specification relevant to the
second order sampling properties of correlograms.

\subsection{Examples of higher-order white noise signals}

To clarify the general white noise sequence $\epsilon[\cdot]$ defined
in the previous section, we give now some explicit examples. The ultimate
example of a white noise signal is of course the Gaussian process
with no memory. This is because the central limit theorem shows that
most processes under general conditions will tend to Gaussian, and
the no-memory aspect implies that the signal is white. The first four
cumulants of a complex Gaussian random variable are\begin{equation}
\left.\begin{array}{l}
\mu=\mu_{g}\\
\sigma^{2}=\sigma_{g}^{2}\\
s^{2}=0\\
\kappa_{3}=0\\
\kappa_{4}=0\end{array}\right\} \quad\textnormal{complex Gaussian white noise}\;\epsilon\sim\textrm{N}(\mu_{g},\sigma_{g}^{2})\end{equation}
Gaussian white noise is, however, not the only type of white noise.
Other kinds of white noise are based on non-Gaussian distributions.
Such situations come about in practice if, for example, the values
of the physical quantity being measured are not continuous as assumed
in the Gaussian case. Examples of this are Poissonian and Bernoullian
white noise.

A Poissonian white noise model is appropriate when the signal values
are limited to non-negative integers as, for instance, when the signal
is a series of independent count values. The first four cumulants
of the real Poissonian distribution, $\textrm{Po}(\lambda)$, which
specify the Poissonian white noise sequence, are\begin{equation}
\left.\begin{array}{l}
\mu=\lambda\\
\sigma^{2}=\lambda\\
s^{2}=\lambda,\;\theta_{2}=0\\
\kappa_{3}=\lambda\\
\kappa_{4}=\lambda\end{array}\right\} \quad\textnormal{real Poissonian white noise}\;\epsilon\sim\textrm{Po}(\lambda)\end{equation}
Notice how the third and fourth order cumulants $\kappa_{3}$ and
$\kappa_{4}$ are nonzero in contrast to Gaussian white noise. The
reality of the signal comes from the fact that $s^{2}=\sigma^{2}$
and $\theta_{2}=0$.

The Bernoullian white noise model is appropriate when the signal values
are binary, for instance, a sequence of independent yes-no decisions.
We take the range of a Bernoulli random variable to be $\{0,1\}$
and the probability of getting a $1$ is $p$ (so the probability
of getting a $0$ is $1-p$). The first four cumulants of the Bernoulli
distribution, \textbf{$\textrm{Be}(p)$,} are\begin{equation}
\left.\begin{array}{l}
\mu=p\\
\sigma^{2}=-p^{2}+p\\
s^{2}=\sigma^{2},\;\theta_{2}=0\\
\kappa_{3}=2p^{3}-3p^{2}+p\\
\kappa_{4}=-6p^{4}+12p^{3}-7p^{2}+p\end{array}\right\} \quad\textnormal{real Bernoullian white noise}\;\epsilon\sim\textrm{Be}(p)\end{equation}
Again, the cumulants of order greater than 2 are not zero as opposed
to Gaussian white noise and the signal is real because the two second-order
cumulants are equal. The estimation of ACF of one-bit sequences and
their sampling properties was discussed in \cite{Weinreb63}.

A feature of both the Poissonian and Bernoullian noise is that they
inherently have non-zero means. Ultimately, for large enough mean
values both the Poisson and Bernoulli distributions can be shown to
tend to Gaussian. But for finite mean values their cumulants are clearly
incompatible with their Gaussian counterparts.

\section{Sampling properties of various correlograms for general white noise}

There are many different types of correlograms. These can be subdivided,
see \cite{Bendat00}, according to whether the correlograms 1) explicitly
involve an attempt to remove the mean of the signal, 2) correlate
two distinct signals or the signal with itself, and 3) normalisation
based on the data samples is applied. The first dichotomy can also
be seen as whether the intention is to estimate correlation functions
or covariance functions. The second dichotomy is the distinction between
cross- and auto-correlation respectively. The last one includes correlation
functions which are normalised by an estimate of the covariance of
the sequence and are sometimes known as coherence functions.

We will now look in detail at various correlograms which are either
estimates of autocorrelation sequences or of autocovariance sequences.
In particular we will examine the classical correlogram, the fixed-length
correlogram, and the classical correlogram in which the sample mean
is subtracted. Most standard treatments allow the correlograms to
be windowed arbitrarily according to a lag window $w[l]$. We however
use a slightly different quantity for weighting the correlogram lag
estimates which we will simply call the weight function $\mathcal{W}[l]$
which includes all normalisation factors. It is related to the lag
window, which for $w[l]=1$ gives an unbiased correlogram estimate,
as $\mathcal{W}[l]=w[l]/(N-|l|)$. As we allow the signal to be complex
we use accordingly complexified correlograms throughout. As for the
signal to be analysed, we will assume that it consists of a finite
number, $N$, of samples of general white noise, $\epsilon[\cdot]$,
as defined in the previous section.

Our goal in this section is to derive the sampling properties of these
correlograms up to second-order, which fundamentally are the expectation
and the covariance of the estimators. We also provide some related
sampling properties, such as the bias, the mean square error (MSE),
asymptotic MSE for large number of samples.

\subsection{\label{sub:Lag-win-ACF-est}Lag window autocorrelation estimator}

We start by considering the most fundamental and simplest of all correlograms,
the \emph{lag window correlogram} $\widehat{R}_{Z}^{(lw)}$; see \cite{Percival93a}.
It was popularised by Blackman and Tukey \cite{Blackman58}. Although
it was originally expressed for real signals, it is easily generalised
to complex signals by requiring it to be a hermitian form. Thus, for
non-negative lags, we define the complex lag windowed correlogram
as\begin{equation}
\widehat{R}_{Z}^{(lw)}[l]:=\mathcal{W}[l]{\displaystyle \sum_{k=1}^{N-l}}Z^{*}[k]Z[k+l]\quad l\ge0\label{eq:DefRc}\end{equation}
and the negative lags are just the complex conjugate of the correlogram
with the opposite (positive) sign\begin{equation}
\widehat{R}_{Z}^{(lw)}[l]:=\left(\widehat{R}_{Z}^{(lw)}[|l|]\right)^{*}\quad l<0\end{equation}

Usually $\widehat{R}_{Z}^{(lw)}$ is used as an autocovariance estimator.
This implies that the signal being analysed, $Z$, is assumed to have
a zero mean, i.e. $E\{ Z\}=0$. However, in what follows, we will
not assume that the mean is zero, in other words, we see $\widehat{R}_{Z}^{(lw)}$
as an autocorrelation estimator. Whether the data samples have a non-zero
mean intentionally or through an oversight is irrelevant. If the mean
is not known apriori and one wishes to remove it the usual method
is to estimate the mean and subtract it from the data before estimating
the correlogram. In this paper we regard this process as a distinct
correlogram and the most commonly used example, the lag windowed autocovariance
estimator, will be treated in section \ref{sub:ACVF-est}.

\subsubsection{Expectation and bias}

The expectation of $\widehat{R}_{\epsilon}^{(lw)}[l]$ is\begin{align}
E\left\{ \widehat{R}_{\epsilon}^{(lw)}[l]\right\}  & =\mathcal{W}[l]{\displaystyle \sum_{k=1}^{N-l}}\left(E\left\{ \epsilon^{'*}[k]\epsilon^{'}[k+l]\right\} +\mu E\left\{ \epsilon^{'*}[k]\right\} +\mu^{*}E\left\{ \epsilon^{'}[k+l]\right\} +|\mu|^{2}\right)=\nonumber \\
 & =\mathcal{W}[l]{\displaystyle \sum_{k=1}^{N-l}}\left(E\left\{ \epsilon^{'*}[k]\epsilon^{'}[k+l]\right\} +\mu E\left\{ \epsilon^{'*}[k]\right\} +\mu^{*}E\left\{ \epsilon^{'}[k+l]\right\} \right)+(N-l)\mathcal{W}[l]|\mu|^{2}=\nonumber \\
 & =\mathcal{W}[l]{\displaystyle \sum_{k=1}^{N-l}}\sigma^{2}\delta_{0l}+(N-l)\mathcal{W}[l]|\mu|^{2}=\nonumber \\
 & =(N-l)\mathcal{W}[l](\sigma^{2}\delta_{0l}+|\mu|^{2})\end{align}
The bias is just the difference between the expectation and the true,
population value. In this case it is therefore, for all $l$,\begin{align}
\mathrm{Bias}\left\{ \widehat{R}_{\epsilon}^{(lw)}[l]\right\}  & =E\left\{ \widehat{R}_{\epsilon}^{(lw)}[l]\right\} -R_{\epsilon\epsilon}[l]=\nonumber \\
 & =(N-l)\mathcal{W}[l](\sigma^{2}\delta_{0l}+|\mu|^{2})-(\sigma^{2}\delta_{0,l}+|\mu|^{2})=\nonumber \\
 & =((N-l)\mathcal{W}[l]-1)(\sigma^{2}\delta_{0l}+|\mu|^{2}),\end{align}
that is,

\myMathBox{\begin{equation}
\mathrm{Bias}\left\{ \widehat{R}_{\epsilon}^{(lw)}[l]\right\} =((N-l)\mathcal{W}[l]-1)(\sigma^{2}\delta_{0l}+|\mu|^{2}),\qquad l=-N,...-1,0+1,...,N.\end{equation}

}

\subsubsection{Covariance}

The covariance of a correlogram is just the covariance of the correlogram
estimate between any two lags, say $l$ and $l'$. The derivation
of the covariance is as follows. For $ll'\ge0$,\begin{align*}
 & \frac{1}{\mathcal{W}[l]\mathcal{W}[l']}\mathrm{Cov}\left\{ \widehat{R}_{\epsilon}^{(lw)}[l],\widehat{R}_{\epsilon}^{(lw)}[l']\right\} =\\
= & {\displaystyle \sum_{k=1}^{N-l}}{\displaystyle \sum_{k'=1}^{N-l'}}\left(\mathrm{Cov}\left\{ Z^{'*}[k]Z^{'}[k+l],Z^{'*}[k']Z^{'}[k'+l']\right\} +\right.\\
 & +|\mu|^{2}\left(\mathrm{Cov}\left\{ Z^{'*}[k],Z^{'*}[k']\right\} +\mathrm{Cov}\left\{ Z^{'}[k+l],Z^{'}[k'+l']\right\} \right)+\\
 & +\mathrm{Cov}\left\{ |\mu|^{2},|\mu|^{2}\right\} +\mathrm{Cov}\left\{ Z^{'*}[k]Z^{'}[k+l],\mu Z^{'*}[k']\right\} +\mathrm{Cov}\left\{ \mu Z^{'*}[k],Z^{'*}[k']Z^{'}[k'+l']\right\} +\\
 & +\mathrm{Cov}\left\{ Z^{'*}[k]Z^{'}[k+l],\mu^{*}Z^{'}[k'+l']\right\} +\mathrm{Cov}\left\{ \mu^{*}Z^{'}[k+l],Z^{'*}[k']Z^{'}[k'+l']\right\} +\\
 & \left.+\mathrm{Cov}\left\{ \mu Z^{'*}[k],\mu^{*}Z^{'}[k'+l']\right\} +\mathrm{Cov}\left\{ \mu^{*}Z^{'}[k+l],\mu Z^{'*}[k']\right\} \right)\\
= & {\displaystyle \sum_{k=1}^{N-|l|}}{\displaystyle \sum_{k'=1}^{N-|l'|}}\left(\mathrm{Cov}\left\{ \epsilon^{'*}[k]\epsilon^{'}[k+l],\epsilon^{'*}[k']\epsilon^{'}[k'+l']\right\} +\right.\\
 & +|\mu|^{2}\left(\mathrm{Cov}\left\{ \epsilon^{'*}[k],\epsilon^{'*}[k']\right\} +\mathrm{Cov}\left\{ \epsilon^{'}[k+l],\epsilon^{'}[k'+l']\right\} \right)+\\
 & +0+\mu E\left\{ \epsilon^{'}[k]\epsilon^{'*}[k+l]\epsilon^{'*}[k']\right\} +\mu^{*}E\left\{ \epsilon^{'}[k]\epsilon^{'*}[k']\epsilon^{'}[k'+l']\right\} +\\
 & +\mu^{*}E\left\{ \epsilon^{'}[k]\epsilon^{'*}[k+l]\epsilon^{'}[k'+l']\right\} +\mu E\left\{ \epsilon^{'*}[k+l]\epsilon^{'*}[k']\epsilon^{'}[k'+l']\right\} +\\
 & \left.+(\mu^{*})^{2}\mathrm{Cov}\left\{ \epsilon^{'*}[k],\epsilon^{'}[k'+l']\right\} +\mu^{2}\mathrm{Cov}\left\{ \epsilon^{'}[k+l],\epsilon^{'*}[k']\right\} \right)\\
= & {\displaystyle \sum_{k=1}^{N-|l|}}{\displaystyle \sum_{k'=1}^{N-|l'|}}\left((\mu_{4}-2\sigma^{4}-|m_{2}|^{2})\delta_{kk'}\delta_{0l}\delta_{0l'}+\sigma^{4}\delta_{kk'}\delta_{ll'}+s^{4}\delta_{k,k'}\delta_{0,l}\delta_{0,l'}+|\mu|^{2}\left(\sigma^{2}\delta_{kk'}+\sigma^{2}\delta_{k+l,k'+l'}\right)+\right.\\
 & \left.+\mu\kappa_{3}^{*}\delta_{kk'}\delta_{0l}+\mu^{*}\kappa_{3}\delta_{kk'}\delta_{0l'}+\mu^{*}\kappa_{3}\delta_{0l}\delta_{k-k',l'}+\mu\kappa_{3}^{*}\delta_{0l'}\delta_{k'-k,l}+(\mu^{*})^{2}m_{2}\delta_{k,k'+l'}+\mu^{2}m_{2}^{*}\delta_{k',k+l}\right)\\
= & {\displaystyle \sum_{k=1}^{N-|l|}}{\displaystyle \sum_{k'=1}^{N-|l'|}}\left((\kappa_{4}+s^{4})\delta_{kk'}\delta_{0l}\delta_{0l'}+(\mu\kappa_{3}^{*}\delta_{kk'}+\mu^{*}\kappa_{3}\delta_{k,k'+l'})\delta_{0l}+(\mu^{*}\kappa_{3}\delta_{kk'}+\mu\kappa_{3}^{*}\delta_{k',k+l})\delta_{0l'}\right.\\
 & \left.+\sigma^{4}\delta_{kk'}\delta_{ll'}+|\mu|^{2}\sigma^{2}\left(\delta_{kk'}+\delta_{k+l,k'+l'}\right)+(\mu^{*})^{2}m_{2}\delta_{k,k'+l'}+\mu^{2}m_{2}^{*}\delta_{k',k+l}\right)\\
= & \delta_{0l}\delta_{0l'}(\kappa_{4}+s^{4})N+\delta_{0l}(\mu\kappa_{3}^{*}(N-l')+\mu^{*}\kappa_{3}(N-l'))+\delta_{0l'}(\mu^{*}\kappa_{3}(N-l)+\mu\kappa_{3}^{*}(N-l))+\delta_{ll'}\sigma^{4}(N-l)+\\
 & +|\mu|^{2}\sigma^{2}\left(N-\max(l,l')+N-\max(l,l')\right)+(\mu^{*})^{2}m_{2}(N-\min(l+l',N))+\mu^{2}m_{2}^{*}(N-\min(l+l',N))\\
= & \delta_{0l}\delta_{0l'}(\kappa_{4}+s^{4})N+\delta_{0l}(\mu\kappa_{3}^{*}+\mu^{*}\kappa_{3})(N-l')+\delta_{0l'}(\mu^{*}\kappa_{3}+\mu\kappa_{3}^{*})(N-l)+\delta_{ll'}\sigma^{4}(N-l)+\\
 & +2|\mu|^{2}\sigma^{2}\left(N-\max(l,l')\right)+((\mu^{*})^{2}m_{2}+\mu^{2}m_{2}^{*})(N-\min(l+l',N))=\\
= & \delta_{0l}\delta_{0l'}(\kappa_{4}+s^{4})N+2\delta_{0l}\Re(\mu\kappa_{3}^{*})(N-l')+2\delta_{0l'}\Re(\mu\kappa_{3}^{*})(N-l)+\delta_{ll'}\sigma^{4}(N-l)+\\
 & +2|\mu|^{2}\sigma^{2}\left(N-\max(l,l')\right)+2\Re(\mu^{2}m_{2}^{*})(N-\min(l+l',N))\end{align*}
where we have used the summation formulas (\ref{eq:LWsum_diag}),
(\ref{eq:LWsum_offdiag}), (\ref{eq:LWsum_offdiagprim}), (\ref{eq:LWsum_2offdiag}),
and (\ref{eq:LWsum_2xUnity}). Here, $\Re(\cdot)$ is the real-part
operator, $\min(a,b)$ is equal to the argument, either $a$ or $b$,
which is less than or equal to the other argument, and conversely
$\max(a,b)$ is equal to the argument which is more than or equal
to the other argument.

From the above derivation we have thus found that the covariance is

\myMathBox{\begin{multline}
\frac{1}{\mathcal{W}[l]\mathcal{W}[l']}\mathrm{Cov}\left\{ \widehat{R}_{\epsilon}^{(lw)}[l],\widehat{R}_{\epsilon}^{(lw)}[l']\right\} =\delta_{0l}\delta_{0l'}(\kappa_{4}+s^{4})N+2\delta_{0l}\Re(\mu\kappa_{3}^{*})(N-|l'|)+2\delta_{0l'}\Re(\mu\kappa_{3}^{*})(N-|l|)+\\
+\delta_{ll'}\sigma^{4}(N-|l|)+2|\mu|^{2}\sigma^{2}\left(N-\max(|l|,|l'|)\right)+2\Re(\mu^{2}m_{2}^{*})(N-\min(|l|+|l'|,N)),\\
ll'\ge0\end{multline}

}

For $ll'\le0$,\begin{align*}
 & \frac{1}{\mathcal{W}[l]\mathcal{W}[l']}\mathrm{Cov}\left\{ \widehat{R}_{\epsilon}^{(lw)}[l],\widehat{R}_{\epsilon}^{(lw)}[l']\right\} =\\
= & {\displaystyle \sum_{k=1}^{N-l}}{\displaystyle \sum_{k'=1}^{N-l'}}(\mathrm{Cov}\left\{ Z^{'}[k]Z^{'*}[k+l],Z^{'*}[k']Z^{'}[k'+l']\right\} +(\mu^{*})^{2}\mathrm{Cov}\left\{ Z^{'}[k],Z^{'*}[k']\right\} +\mu^{2}\mathrm{Cov}\left\{ Z^{'*}[k+l],Z^{'}[k'+l']\right\} +\\
 & +\mathrm{Cov}\left\{ |\mu|^{2},|\mu|^{2}\right\} +\mathrm{Cov}\left\{ Z^{'}[k]Z^{'*}[k+l],\mu Z^{'*}[k']\right\} +\mathrm{Cov}\left\{ \mu^{*}Z^{'}[k],Z^{'*}[k']Z^{'}[k'+l']\right\} +\\
 & +\mathrm{Cov}\left\{ Z^{'}[k]Z^{'*}[k+l],\mu^{*}Z^{'}[k'+l']\right\} +\mathrm{Cov}\left\{ \mu Z^{'*}[k+l],Z^{'*}[k']Z^{'}[k'+l']\right\} +\mathrm{Cov}\left\{ \mu^{*}Z^{'}[k],\mu^{*}Z^{'}[k'+l']\right\} +\\
 & +\mathrm{Cov}\left\{ \mu Z^{'*}[k+l],\mu Z^{'*}[k']\right\} )\\
= & {\displaystyle \sum_{k=1}^{N-|l|}}{\displaystyle \sum_{k'=1}^{N-|l'|}}(\mathrm{Cov}\left\{ \epsilon^{'*}[k]\epsilon^{'}[k+l],\epsilon^{'*}[k']\epsilon^{'}[k'+l']\right\} +(\mu^{*})^{2}\mathrm{Cov}\left\{ \epsilon^{'}[k],\epsilon^{'*}[k']\right\} +\mu^{2}\mathrm{Cov}\left\{ \epsilon^{'*}[k+l],\epsilon^{'}[k'+l']\right\} +\\
 & +0+\mu E\left\{ \epsilon^{'*}[k]\epsilon^{'}[k+l]\epsilon^{'*}[k']\right\} +\mu E\left\{ \epsilon^{'*}[k]\epsilon^{'*}[k']\epsilon^{'}[k'+l']\right\} +\\
 & +\mu^{*}E\left\{ \epsilon^{'*}[k]\epsilon^{'}[k+l]\epsilon^{'}[k'+l']\right\} +\mu^{*}E\left\{ \epsilon^{'}[k+l]\epsilon^{'*}[k']\epsilon^{'}[k'+l']\right\} +\\
 & +|\mu|^{2}\mathrm{Cov}\left\{ \epsilon^{'}[k],\epsilon^{'}[k'+l']\right\} +|\mu|^{2}\mathrm{Cov}\left\{ \epsilon^{'*}[k+l],\epsilon^{'*}[k']\right\} )\\
= & {\displaystyle \sum_{k=1}^{N-|l|}}{\displaystyle \sum_{k'=1}^{N-|l'|}}((\mu_{4}-2\sigma^{4}-|m_{2}|^{2})\delta_{kk'}\delta_{0l}\delta_{0l'}+s^{4}\delta_{kk'}\delta_{ll'}+\sigma^{4}\delta_{k,k'}\delta_{0,l}\delta_{0,l'}+(\mu^{*})^{2}m_{2}^{*}\delta_{kk'}+\mu^{2}m_{2}\delta_{k+l,k'+l'}+\\
 & +\mu\kappa_{3}^{*}\delta_{kk'}\delta_{0l}+\mu\kappa_{3}^{*}\delta_{kk'}\delta_{0l'}+\mu^{*}\kappa_{3}\delta_{0l}\delta_{k-k',l'}+\mu^{*}\kappa_{3}\delta_{0l'}\delta_{k'-k,l}+\\
 & +|\mu|^{2}\sigma^{2}\delta_{k,k'+l'}+|\mu|^{2}\sigma^{2}\delta_{k',k+l})\\
= & {\displaystyle \sum_{k=1}^{N-|l|}}{\displaystyle \sum_{k'=1}^{N-|l'|}}((\kappa_{4}+\sigma^{4})\delta_{kk'}\delta_{0l}\delta_{0l'}+(\mu\kappa_{3}^{*}\delta_{kk'}+\mu^{*}\kappa_{3}\delta_{k,k'+l'})\delta_{0l}+(\mu\kappa_{3}^{*}\delta_{kk'}+\mu^{*}\kappa_{3}\delta_{k',k+l})\delta_{0l'}+s^{4}\delta_{kk'}\delta_{ll'}+\\
 & +|\mu|^{2}\sigma^{2}\left(\delta_{k,k'+l'}+\delta_{k',k+l}\right)+(\mu^{*})^{2}m_{2}^{*}\delta_{k,k'}+\mu^{2}m_{2}\delta_{k+l,k'+l'})\\
= & \delta_{0l}\delta_{0l'}(\kappa_{4}+\sigma^{4})N+\delta_{0l}(\mu\kappa_{3}^{*}(N-l')+\mu^{*}\kappa_{3}(N-l'))+\delta_{0l'}(\mu\kappa_{3}^{*}(N-l)+\mu^{*}\kappa_{3}(N-l))+\delta_{ll'}s^{4}(N-l)+\\
 & +|\mu|^{2}\sigma^{2}\left(N-\max(l,l')+N-\max(l,l')\right)+(\mu^{*})^{2}m_{2}^{*}(N-\min(l+l',N))+\mu^{2}m_{2}(N-\min(l+l',N))\\
= & \delta_{0l}\delta_{0l'}(\kappa_{4}+\sigma^{4})N+\delta_{0l}(\mu\kappa_{3}^{*}+\mu^{*}\kappa_{3})(N-l')+\delta_{0l'}(\mu^{*}\kappa_{3}+\mu\kappa_{3}^{*})(N-l)+\delta_{ll'}s^{4}(N-l)+\\
 & +2|\mu|^{2}\sigma^{2}\left(N-\max(l,l')\right)+((\mu^{*})^{2}m_{2}^{*}+\mu^{2}m_{2})(N-\min(l+l',N))=\\
= & \delta_{0l}\delta_{0l'}(\kappa_{4}+\sigma^{4})N+2\delta_{0l}\Re(\mu\kappa_{3}^{*})(N-l')+2\delta_{0l'}\Re(\mu\kappa_{3}^{*})(N-l)+\delta_{ll'}s^{4}(N-l)+\\
 & +2|\mu|^{2}\sigma^{2}\left(N-\max(l,l')\right)+2\Re(\mu^{2}m_{2})(N-\min(l+l',N))\end{align*}
so\begin{align}
\frac{1}{\mathcal{W}[l]\mathcal{W}[l']}\mathrm{Cov}\left\{ \widehat{R}_{\epsilon}^{(lw)}[l],\widehat{R}_{\epsilon}^{(lw)}[l']\right\} = & \delta_{0l}\delta_{0l'}(\kappa_{4}+\sigma^{4})N+2\delta_{0l}\Re(\mu\kappa_{3}^{*})(N-|l'|)+2\delta_{0l'}\Re(\mu\kappa_{3}^{*})(N-|l|)+\nonumber \\
 & +\delta_{ll'}s^{4}(N-|l|)+2|\mu|^{2}\sigma^{2}\left(N-\max(|l|,|l'|)\right)\nonumber \\
 & +2\Re(\mu^{2}m_{2})(N-\min(|l|+|l'|,N)),\qquad ll'\le0\end{align}

\subsubsection{Mean square error}

Now the mean square error (MSE) is the variance ($l=l'$) plus the
bias squared. The derivation of the MSE in this case starts from this
definition and proceeds as follows:\begin{align*}
 & \mathrm{MSE}\left\{ \widehat{R}_{\epsilon\epsilon}^{(lw)}[l]\right\} =\mathrm{Var}\left\{ \widehat{R}_{\epsilon\epsilon}^{(lw)}[l]\right\} +\left|\mathrm{Bias}\left\{ \widehat{R}_{\epsilon\epsilon}^{(lw)}[l]\right\} \right|^{2}=\\
= & \mathcal{W}^{2}[l]\left(\delta_{0l}((\kappa_{4}+s^{4})+4\Re(\mu\mu_{3}^{*}))N+(\sigma^{4}+2|\mu|^{2}\sigma^{2})\left(N-l\right)+2\Re(\mu^{2}m_{2}^{*})(N-\min(2l,N))\right)+\\
 & +|(N-l)\mathcal{W}[l]-1|^{2}(\sigma^{2}\delta_{0l}+|\mu|^{2})^{2}=\\
= & \mathcal{W}^{2}[l]\left(\delta_{0l}((\kappa_{4}+s^{4})+4\Re(\mu\mu_{3}^{*}))N+(\sigma^{4}+2|\mu|^{2}\sigma^{2})\left(N-l\right)+2\Re(\mu^{2}m_{2}^{*})(N-\min(2l,N))\right)+\\
 & +((N-l)^{2}\mathcal{W}^{2}[l]-2(N-l)\mathcal{W}[l]+1)((\sigma^{4}+2\sigma^{2}|\mu|^{2})\delta_{0,l}+|\mu|^{4})=\\
= & \delta_{0l}\left(((\kappa_{4}+s^{4})+4\Re(\mu\mu_{3}^{*}))N\mathcal{W}^{2}[0]+\sigma^{2}(\sigma^{2}+2|\mu|^{2})(N^{2}\mathcal{W}^{2}[0]-2N\mathcal{W}[0]+1)\right)+\\
 & +\left(\sigma^{2}(\sigma^{2}+2|\mu|^{2})\left(N-l\right)+2\Re(\mu^{2}m_{2}^{*})(N-\min(2l,N))+|\mu|^{4}(N-l)^{2}\right)\mathcal{W}^{2}[l]-2|\mu|^{4}(N-l)\mathcal{W}[l]+|\mu|^{4}=\\
= & \delta_{0l}\left(((\kappa_{4}+s^{4})+4\Re(\mu\mu_{3}^{*})+\sigma^{2}(\sigma^{2}+2|\mu|^{2})N)N\mathcal{W}^{2}[0]-2N\sigma^{2}(\sigma^{2}+2|\mu|^{2})\mathcal{W}[0]+\sigma^{2}(\sigma^{2}+2|\mu|^{2}))\right)+\\
 & +\left(\sigma^{2}(\sigma^{2}+2|\mu|^{2})\left(N-l\right)+2\Re(\mu^{2}m_{2}^{*})(N-\min(2l,N))+|\mu|^{4}(N-l)^{2}\right)\mathcal{W}^{2}[l]-2|\mu|^{4}(N-l)\mathcal{W}[l]+|\mu|^{4}\end{align*}

Thus we find that the MSE is

\myMathBox{\begin{align}
\mathrm{MSE}\left\{ \widehat{R}_{\epsilon\epsilon}^{(lw)}[0]\right\} = & \left(\kappa_{4}+s^{4}+\sigma^{4}+2|\mu|^{2}\sigma^{2}+4\Re(\mu\kappa_{3}^{*})+2\Re(\mu^{2}m_{2}^{*})+\left(\sigma^{2}+|\mu|^{2}\right)^{2}N\right)N\mathcal{W}^{2}[0]+\nonumber \\
 & -2N\left(\sigma^{2}+|\mu|^{2}\right)^{2}\mathcal{W}[0]+\left(\sigma^{2}+|\mu|^{2}\right)^{2}\\
\mathrm{MSE}\left\{ \widehat{R}_{\epsilon\epsilon}^{(lw)}[l\neq0]\right\} = & \left(\sigma^{2}(\sigma^{2}+2|\mu|^{2})\left(N-l\right)+2\Re(\mu^{2}m_{2}^{*})(N-\min(2l,N))+|\mu|^{4}(N-l)^{2}\right)\mathcal{W}^{2}[l]+\nonumber \\
 & -2|\mu|^{4}(N-l)\mathcal{W}[l]+|\mu|^{4}\end{align}

}

Asymptotically, i.e. $N\rightarrow\infty$, keeping only the leading
terms of in $N$ or $l$ for each coefficient of each power of $\mathcal{W}[l]$,
we find\begin{align}
\mathrm{MSE}\left\{ \widehat{R}_{\epsilon}^{(lw)}[0]\right\}  & =(|\mu|^{2}+\sigma^{2})^{2}N^{2}\mathcal{W}^{2}[0]-2N(\sigma^{2}+|\mu|^{2})^{2}\mathcal{W}[0]+(\sigma^{2}+|\mu|^{2})^{2}\\
\mathrm{MSE}\left\{ \widehat{R}_{\epsilon}^{(lw)}[l\neq0]\right\}  & =|\mu|^{4}(N-l)^{2}\mathcal{W}^{2}[l]-2|\mu|^{4}(N-l)\mathcal{W}[l]+|\mu|^{4}\end{align}
where we assume that $\mu\neq0$.

\subsection{\label{sub:Fix-len-ACF-est}Fixed-length summation autocorrelation
estimator}

As an example of a correlogram that does not have form of the classic
lag windowed correlogram, as given by equation (\ref{eq:DefRc}),
we present what we will call the \emph{fixed-length summation correlogram}
denoted $\widehat{R}_{Z}^{(fl)}$. It is defined as

\begin{equation}
\widehat{R}_{Z}^{(fl)}[l]:=\mathcal{W}[l]{\displaystyle \sum_{k=1}^{L}}Z^{*}[k]Z[k+l]\quad0\le l\le N-L=M,\end{equation}
This estimator is defined in \cite{Bendat00}, equation (8.96). It
is implemented on the DWP electron counts correlator experiment \cite{Buckley2000}
onboard the CLUSTER-II space mission. Its defining characteristic
is that the same, fixed number of terms are summed over for all lags.
This in contrast with the classic correlogram for which the number
of terms decrements with increasing lag. This implies that the lags
of the classical correlogram all have different statistics, which
is an unwanted property. With a fixed length summation however, one
might hope that each of the estimated lags will have the same statistics
giving more equitable lag estimates. As we will now see, this is in
fact true but only when the mean is zero.

\subsubsection{Expectation and bias}

The bias is for a general signal \begin{align*}
E\left\{ \widehat{R}_{Z}^{(fl)}[l]\right\}  & =\mathcal{W}[l]{\displaystyle \sum_{k=1}^{L}}E\left\{ Z^{*}[k]Z[k+l]\right\} =\\
 & =\mathcal{W}[l]{\displaystyle \sum_{k=1}^{L}}R_{Z}[l]=\\
 & =\mathcal{W}[l]R_{Z}[l]{\displaystyle \sum_{k=1}^{L}}1=\\
 & =L\mathcal{W}[l]R_{Z}[l]\end{align*}
The bias is then\begin{align*}
\mathrm{Bias}\left\{ \widehat{R}_{Z}^{(fl)}[l]\right\}  & =E\left\{ \widehat{R}_{Z}^{(fl)}[l]\right\} -R_{Z}[l]=\\
 & =L\mathcal{W}[l]R_{Z}[l]-R_{Z}[l]=\\
 & =\left(L\mathcal{W}[l]-1\right)R_{Z}[l]\end{align*}
Thus with a window choice of $\mathcal{W}[l]=1/L$ the fixed-length
correlogram is unbiased irrespective of the signal. For the special
case of the higher-order white noise signal, $\epsilon$, with which
we are mainly interested in here, the bias is therefore

\myMathBox{\begin{equation}
\mathrm{Bias}\left\{ \widehat{R}_{\epsilon}^{(fl)}[l]\right\} =\left(L\mathcal{W}[l]-1\right)\left(\sigma^{2}\delta_{0,l}+|\mu|^{2}\right)\end{equation}

}

\subsubsection{Covariance}

The covariance of the fixed-length correlogram can be derived accordingly,

\begin{align*}
 & \frac{1}{\mathcal{W}[l]\mathcal{W}[l']}\mathrm{Cov}\left\{ \widehat{R}_{\epsilon}^{(fl)}[l],\widehat{R}_{\epsilon}^{(fl)}[l']\right\} =\\
= & {\displaystyle \sum_{k=1}^{L}}{\displaystyle \sum_{k'=1}^{L}}\left(\mathrm{Cov}\left\{ Z^{'*}[k]Z^{'}[k+l],Z^{'*}[k']Z^{'}[k'+l']\right\} \right.+\\
 & \hphantom{{\displaystyle \sum_{k=1}^{L}}{\displaystyle \sum_{k'=1}^{L}}\left(\right.}+|\mu|^{2}\left(\mathrm{Cov}\left\{ Z^{'*}[k],Z^{'*}[k']\right\} +\mathrm{Cov}\left\{ Z^{'}[k+l],Z^{'}[k'+l']\right\} \right)+\mathrm{Cov}\left\{ |\mu|^{2},|\mu|^{2}\right\} +\\
 & \hphantom{{\displaystyle \sum_{k=1}^{L}}{\displaystyle \sum_{k'=1}^{L}}\left(\right.}+\mathrm{Cov}\left\{ Z^{'*}[k]Z^{'}[k+l],\mu Z^{'*}[k']\right\} +\mathrm{Cov}\left\{ \mu Z^{'*}[k],Z^{'*}[k']Z^{'}[k'+l']\right\} +\\
 & \hphantom{{\displaystyle \sum_{k=1}^{L}}{\displaystyle \sum_{k'=1}^{L}}\left(\right.}+\mathrm{Cov}\left\{ Z^{'*}[k]Z^{'}[k+l],\mu^{*}Z^{'}[k'+l']\right\} +\mathrm{Cov}\left\{ \mu^{*}Z^{'}[k+l],Z^{'*}[k']Z^{'}[k'+l']\right\} +\\
 & \hphantom{{\displaystyle \sum_{k=1}^{L}}{\displaystyle \sum_{k'=1}^{L}}\left(\right.}\left.+\mathrm{Cov}\left\{ \mu Z^{'*}[k],\mu^{*}Z^{'}[k'+l']\right\} +\mathrm{Cov}\left\{ \mu^{*}Z^{'}[k+l],\mu Z^{'*}[k']\right\} \right)=\\
= & {\displaystyle \sum_{k=1}^{L}}{\displaystyle \sum_{k'=1}^{L}}\left(\mathrm{Cov}\left\{ \epsilon^{'*}[k]\epsilon^{'}[k+l],\epsilon^{'*}[k']\epsilon^{'}[k'+l']\right\} \right.+\\
 & \hphantom{{\displaystyle \sum_{k=1}^{L}}{\displaystyle \sum_{k'=1}^{L}}\left(\right.}+(\mu^{*})^{2}\mathrm{Cov}\left\{ \epsilon^{'}[k],\epsilon^{'*}[k']\right\} +\mu^{2}\mathrm{Cov}\left\{ \epsilon^{'*}[k+l],\epsilon^{'}[k'+l']\right\} +0+\\
 & \hphantom{{\displaystyle \sum_{k=1}^{L}}{\displaystyle \sum_{k'=1}^{L}}\left(\right.}+\mu E\left\{ \epsilon^{'*}[k]\epsilon^{'}[k+l]\epsilon^{'*}[k']\right\} +\mu E\left\{ \epsilon^{'*}[k]\epsilon^{'*}[k']\epsilon^{'}[k'+l']\right\} +\\
 & \hphantom{{\displaystyle \sum_{k=1}^{L}}{\displaystyle \sum_{k'=1}^{L}}\left(\right.}+\mu^{*}E\left\{ \epsilon^{'*}[k]\epsilon^{'}[k+l]\epsilon^{'}[k'+l']\right\} +\mu^{*}E\left\{ \epsilon^{'}[k+l]\epsilon^{'*}[k']\epsilon^{'}[k'+l']\right\} +\\
 & \hphantom{{\displaystyle \sum_{k=1}^{L}}{\displaystyle \sum_{k'=1}^{L}}\left(\right.}\left.+|\mu|^{2}\mathrm{Cov}\left\{ \epsilon^{'}[k],\epsilon^{'}[k'+l']\right\} +|\mu|^{2}\mathrm{Cov}\left\{ \epsilon^{'*}[k+l],\epsilon^{'*}[k']\right\} \right)=\\
= & {\displaystyle \sum_{k=1}^{L}}{\displaystyle \sum_{k'=1}^{L}}\left((\mu_{4}-2\sigma^{4}-|m_{2}|^{2})\delta_{kk'}\delta_{0l}\delta_{0l'}+s^{4}\delta_{kk'}\delta_{ll'}+\sigma^{4}\delta_{k,k'}\delta_{0,l}\delta_{0,l'}+(\mu^{*})^{2}m_{2}^{*}\delta_{kk'}+\mu^{2}m_{2}\delta_{k+l,k'+l'}\right.+\\
 & \hphantom{{\displaystyle \sum_{k=1}^{L}}{\displaystyle \sum_{k'=1}^{L}}\left(\right.}\left.+\mu\kappa_{3}^{*}\delta_{kk'}\delta_{0l}+\mu\kappa_{3}^{*}\delta_{kk'}\delta_{0l'}+\mu^{*}\kappa_{3}\delta_{0l}\delta_{k-k',l'}+\mu^{*}\kappa_{3}\delta_{0l'}\delta_{k'-k,l}+|\mu|^{2}\sigma^{2}\delta_{k,k'+l'}+|\mu|^{2}\sigma^{2}\delta_{k',k+l}\right)=\\
= & {\displaystyle \sum_{k=1}^{L}}{\displaystyle \sum_{k'=1}^{L}}\left((\kappa_{4}+s^{4})\delta_{kk'}\delta_{0l}\delta_{0l'}+(\mu\kappa_{3}^{*}\delta_{kk'}+\mu^{*}\kappa_{3}\delta_{k,k'+l'})\delta_{0l}+(\mu^{*}\kappa_{3}\delta_{kk'}+\mu\kappa_{3}^{*}\delta_{k',k+l})\delta_{0l'}\right.+\\
 & \hphantom{{\displaystyle \sum_{k=1}^{L}}{\displaystyle \sum_{k'=1}^{L}}\left(\right.}\left.+\sigma^{4}\delta_{kk'}\delta_{ll'}+|\mu|^{2}\sigma^{2}\left(\delta_{k,k'+l'}+\delta_{k',k+l}\right)+(\mu^{*})^{2}m_{2}\delta_{k,k'}+\mu^{2}m_{2}^{*}\delta_{k+l,k'+l'}\right)=\\
= & \delta_{0l}\delta_{0l'}(\kappa_{4}+s^{4})L+\delta_{0l}\left(\mu\kappa_{3}^{*}L+\mu^{*}\kappa_{3}(L-\min(|l'|,L))\right)+\delta_{0l'}\left(\mu^{*}\kappa_{3}L+\mu\kappa_{3}^{*}(L-\min(|l|,L))\right)+\\
 & +\delta_{ll'}\sigma^{4}L+|\mu|^{2}\sigma^{2}\left(L-\min(|l'|,L)+L-\min(|l|,L)\right)+(\mu^{*})^{2}m_{2}L+\mu^{2}m_{2}^{*}\left(L-\min(||l|-|l'||,L)\right)=\\
= & \delta_{0l}\delta_{0l'}(\kappa_{4}+s^{4})L+\delta_{0l}\left(2\Re(\mu\kappa_{3}^{*})L-\mu^{*}\kappa_{3}\min(|l'|,L)\right)+\delta_{0l'}\left(2\Re(\mu\kappa_{3}^{*})L-\mu\kappa_{3}^{*}\min(|l|,L)\right)+\\
 & +\delta_{ll'}\sigma^{4}L+|\mu|^{2}\sigma^{2}\left(2L-\min(|l|,L)-\min(|l'|,L)\right)+2\Re(\mu^{2}m_{2}^{*})L-\mu^{2}m_{2}^{*}\min(||l|-|l'||,L)\end{align*}
where we have used the summation formulas (\ref{eq:FLsum_diag}),
(\ref{eq:FLsum_offdiag}), (\ref{eq:FLsum_offdiagprim}), (\ref{eq:FLsum_2offdiag}),
and (\ref{eq:FLsum_2xUnity}).

So finally we can write the covariance for all $l$ and $l'$ as

\myMathBox{\begin{multline}
\frac{1}{\mathcal{W}[l]\mathcal{W}[l']}\mathrm{Cov}\left\{ \widehat{R}_{\epsilon}^{(fl)}[l],\widehat{R}_{\epsilon}^{(fl)}[l']\right\} =\delta_{0l}\delta_{0l'}(\kappa_{4}+s^{4})L+\delta_{0l}\left(2\Re(\mu\kappa_{3}^{*})L-\mu^{*}\kappa_{3}\min(|l'|,L)\right)+\\
+\delta_{0l'}\left(2\Re(\mu\kappa_{3}^{*})L-\mu\kappa_{3}^{*}\min(|l|,L)\right)+\delta_{ll'}\sigma^{4}L+\\
+|\mu|^{2}\sigma^{2}\left(2L-\min(|l|,L)-\min(|l'|,L)\right)+2\Re(\mu^{2}m_{2}^{*})L-\mu^{2}m_{2}^{*}\min(||l|-|l'||,L)\end{multline}
}

For convenience, we also provide the covariance for the case of real-valued
variates\begin{align}
\frac{1}{\mathcal{W}[l]\mathcal{W}[l']}\mathrm{Cov}\left\{ \widehat{R}_{\epsilon}^{(fl)}[l],\widehat{R}_{\epsilon}^{(fl)}[l']\right\} = & \delta_{0l}\delta_{0l'}(\kappa_{4}+s^{4})L+\delta_{0l}\mu\kappa_{3}(2L-\min(|l'|,L))+\nonumber \\
 & +\delta_{0l'}\mu\kappa_{3}(2L-\min(|l|,L))+\delta_{|l|,|l'|}\sigma^{4}L+\nonumber \\
 & +\mu^{2}\sigma^{2}\left(4L-\min(|l|,L)-\min(|l'|,L)-\min(||l|-|l'||,L)\right)\end{align}
and if in addition $|l'|\le|l|\le L$ , then \begin{align}
\frac{1}{\mathcal{W}[l]\mathcal{W}[l']}\mathrm{Cov}\left\{ \widehat{R}_{\epsilon}^{(fl)}[l],\widehat{R}_{\epsilon}^{(fl)}[l']\right\} = & \delta_{0l}\delta_{0l'}(\kappa_{4}+s^{4})L+\delta_{0l}2\mu\kappa_{3}(L-|l'|/2)+\delta_{0l'}2\mu\kappa_{3}(L-|l|/2)+\nonumber \\
 & +\delta_{ll'}\sigma^{4}L+4\mu^{2}\sigma^{2}(L-|l|)/2)\end{align}

The results shown above are expressed for both zero and nonzero lags.
Formulas can also be given which are explicit in the zero and nonzero
lags. They are as follows:\begin{align}
\mathrm{Var}\left\{ \widehat{R}_{\epsilon}^{(fl)}[0]\right\}  & =\mathcal{W}^{2}[0]L\left(\kappa_{4}+s^{4}+\sigma^{4}+2|\mu|^{2}\sigma^{2}+2\Re(\mu^{2}m_{2}^{*})+4\Re(\mu\kappa_{3}^{*})\right)\\
\mathrm{Var}\left\{ \widehat{R}_{\epsilon}^{(fl)}[l\neq0]\right\}  & =\mathcal{W}^{2}[l]\left(L\sigma^{4}+2|\mu|^{2}\sigma^{2}(L-\min(|l|,L))+2L\Re(\mu^{2}m_{2}^{*})\right)\end{align}
\begin{equation}
\frac{1}{\mathcal{W}[0]\mathcal{W}[l]}\mathrm{Cov}\left\{ \widehat{R}_{\epsilon}^{(fl)}[0],\widehat{R}_{\epsilon}^{(fl)}[l]\right\} =2L\left(\Re(\mu\kappa_{3}^{*})+\Re(\mu^{2}m_{2}^{*})+|\mu|^{2}\sigma^{2}\right)-\left(|\mu|^{2}\sigma^{2}+\mu^{*}\kappa_{3}+\mu^{2}m_{2}^{*}\right)\min(|l|,L),\quad l\neq0\end{equation}
\begin{multline*}
\frac{1}{\mathcal{W}[l]\mathcal{W}[l']}\mathrm{Cov}\left\{ \widehat{R}_{\epsilon}^{(fl)}[l],\widehat{R}_{\epsilon}^{(fl)}[l']\right\} =|\mu|^{2}\sigma^{2}(2L-\min(|l|,L)-\min(|l'|,L))\\
+2\Re(\mu^{2}m_{2}^{*})L-\mu^{2}m_{2}^{*}\min(||l|-|l'||,L),\quad l\neq0,l'\neq0\end{multline*}

\subsubsection{Mean square error}

The mean square error of the fixed length correlogram for complex
white noise with arbitrary mean can now be derived. The derivation
proceeds from the variance and bias results derived above as follows

\begin{align*}
 & \mathrm{MSE}\left\{ \widehat{R}_{\epsilon}^{(fl)}[l]\right\} =\mathrm{Var}\left\{ \widehat{R}_{\epsilon}^{(fl)}[l]\right\} +\left|\mathrm{Bias}\left\{ \widehat{R}_{\epsilon}^{(fl)}[l]\right\} \right|^{2}=\\
= & \mathcal{W}^{2}[l]\left(\delta_{0l}(\kappa_{4}+s^{4})L+\delta_{0l}4\Re(\mu\kappa_{3}^{*})L+\sigma^{4}L+2|\mu|^{2}\sigma^{2}\left(L-\min(|l|,L)\right)+2\Re(\mu^{2}m_{2}^{*})L\right)\\
 & +|L\mathcal{W}[l]-1|^{2}\left(\sigma^{2}\delta_{0l}+|\mu|^{2}\right)^{2}\\
= & \mathcal{W}^{2}[l]\left(\delta_{0l}L(\kappa_{4}+s^{4}+4\Re(\mu\kappa_{3}^{*}))+\sigma^{4}L+2|\mu|^{2}\sigma^{2}\left(L-\min(|l|,L)\right)+2\Re(\mu^{2}m_{2}^{*})L\right)\\
 & +(L^{2}\mathcal{W}^{2}[l]-2L\mathcal{W}[l]+1)\left(\sigma^{4}\delta_{0l}+2\sigma^{2}|\mu|^{2}\delta_{0l}+|\mu|^{4}\right)\\
= & \delta_{0l}\left(\mathcal{W}^{2}[l]L\left(\kappa_{4}+s^{4}+4\Re(\mu\kappa_{3}^{*})\right)+\sigma^{2}(\sigma^{2}+2|\mu|^{2})(L^{2}\mathcal{W}^{2}[l]-2L\mathcal{W}[l]+1)\right)\\
 & +\mathcal{W}^{2}[l]\left(\sigma^{4}L+2|\mu|^{2}\sigma^{2}\left(L-\min(|l|,L)\right)+2\Re(\mu^{2}m_{2}^{*})L\right)+|\mu|^{4}\left(L^{2}\mathcal{W}^{2}[l]-2L\mathcal{W}[l]+1\right)\\
= & \delta_{0l}\left(\mathcal{W}^{2}[l]L\left(\kappa_{4}+s^{4}+4\Re(\mu\kappa_{3}^{*})+L\sigma^{2}(\sigma^{2}+2|\mu|^{2})\right)+\sigma^{2}(\sigma^{2}+2|\mu|^{2})(1-2L\mathcal{W}[l])\right)\\
 & +\mathcal{W}^{2}[l]\left(\sigma^{4}L+2|\mu|^{2}\sigma^{2}\left(L-\min(|l|,L)\right)+2\Re(\mu^{2}m_{2}^{*})L+L^{2}|\mu|^{4}\right)+|\mu|^{4}\left(1-2L\mathcal{W}[l]\right)\end{align*}
Thus, the final formula for the MSE is

\myMathBox{\begin{align}
\mathrm{MSE}\left\{ \widehat{R}_{\epsilon}^{(fl)}[0]\right\} = & L\left(\kappa_{4}+s^{4}+4\Re(\mu\kappa_{3}^{*})+L(\sigma^{2}+|\mu|^{2})^{2}+\sigma^{4}+2|\mu|^{2}\sigma^{2}+2\Re(\mu^{2}m_{2}^{*})\right)\mathcal{W}^{2}[0]\nonumber \\
 & +\left(\sigma^{2}+|\mu|^{2}\right)^{2}\left(1-2L\mathcal{W}[0]\right)\\
\mathrm{MSE}\left\{ \widehat{R}_{\epsilon}^{(fl)}[l\neq0]\right\} = & \left(L\sigma^{4}+2|\mu|^{2}\sigma^{2}\left(L-\min(|l|,L)\right)+2L\Re(\mu^{2}m_{2}^{*})+L^{2}|\mu|^{4}\right)\mathcal{W}^{2}[l]\nonumber \\
 & +|\mu|^{4}\left(1-2L\mathcal{W}[l]\right)\end{align}

}

Asymptotically as $L\rightarrow\infty$ and assuming $\mu\neq0$ the
MSE becomes\begin{align*}
\mathrm{MSE}\left\{ \widehat{R}_{\epsilon}^{(fl)}[0]\right\} = & \left(\sigma^{2}+|\mu|^{2}\right)^{2}\left(L^{2}\mathcal{W}^{2}[0]-2L\mathcal{W}[0]+1\right)\\
\mathrm{MSE}\left\{ \widehat{R}_{\epsilon}^{(fl)}[l\neq0]\right\} = & |\mu|^{4}\left(L^{2}\mathcal{W}^{2}[l]-2L\mathcal{W}[l]+1\right)\end{align*}
 where we have kept leading terms in $L$ for each power of $\mathcal{W}$.
This proves that for the weight window choice $\mathcal{W}[l]=1/L$
the fixed-length correlogram, as an autocorrelation estimate, is statistically
consistent since the MSE is zero at all lags. If instead $\mu=0$
then the asymptotic MSE of the nonzero lags is $L\sigma^{4}|\mathcal{W}[l]|^{2}$.

\subsection{\label{sub:ACVF-est}Lag windowed autocovariance estimator using
sample mean}

The two previous correlograms cannot be used as autocovariance estimators
if the signal has an apriori unknown mean. As this is a common state
of affairs one usually resorts to estimating the mean and subtracting
this estimate from the signal. If the mean estimate is based on the
same data samples as those on which the autocovariance estimate is
to be based it is natural to see this scheme as a distinct estimator.
In fact the mean need not even be explicitly calculated. There are
a variety of autocovariance estimators, see \cite{Marriott54}. Here
we only consider the simplest: the classic correlogram in which the
ordinary sample mean is subtracted from the data samples. We denote
it $\widehat{C}_{Z}^{(lw)}$ and defined it for an arbitrary complex
signal as\begin{equation}
\widehat{C}_{Z}^{(lw)}[l]:=\mathcal{W}[l]\sum(Z_{k}-\overline{Z})(Z_{k+l}-\overline{Z})\end{equation}
where\begin{equation}
\overline{Z}=\frac{1}{N}{\displaystyle \sum_{k=1}^{N}}Z_{k}\label{eq:DefSampMean}\end{equation}
is the sample mean.

\subsubsection{Properties of the sample mean}

In order to derive the sampling properties of $\widehat{C}_{Z}^{(lw)}$,
we will need some of the properties of the sample mean estimator (\ref{eq:DefSampMean}).
These well know properties are that the sample mean estimator is unbiased,

\begin{equation}
E\left\{ \overline{Z}\right\} =E\left\{ \frac{1}{N}{\displaystyle \sum_{k=1}^{N}}Z_{k}\right\} =\frac{1}{N}{\displaystyle \sum_{k=1}^{N}}E\left\{ Z_{k}\right\} =\frac{1}{N}{\displaystyle \sum_{k=1}^{N}}\mu=\mu\end{equation}
that the standard deviation for independent samples goes is proportional
to $1/\sqrt{N}$,\begin{align}
\mathrm{Var}\left\{ \overline{\epsilon}\right\}  & =\frac{1}{N^{2}}{\displaystyle \sum_{k=1}^{N}}{\displaystyle \sum_{k'=1}^{N}}E\left\{ \epsilon[k]\epsilon[k']^{*}\right\} -E\left\{ \overline{\epsilon[k]}\right\} ^{2}=\frac{1}{N^{2}}{\displaystyle \sum_{k=1}^{N}}{\displaystyle \sum_{k'=1}^{N}}(\sigma^{2}\delta_{kk'}+\mu^{2})-\mu^{2}\nonumber \\
 & =\frac{\sigma^{2}}{N^{2}}{\displaystyle \sum_{k=1}^{N}}{\displaystyle \sum_{k'=1}^{N}}\delta_{kk'}+\mu^{2}-\mu^{2}=\frac{\sigma^{2}}{N^{2}}N\nonumber \\
 & =\frac{\sigma^{2}}{N}\end{align}
where we have specialised to the generalised noise signal $\epsilon[k]$.

\subsubsection{Expectation and bias}

The expectation of $\widehat{C}_{Z}^{(lw)}$ is

\begin{equation}
E\left\{ \widehat{C}_{\epsilon;\mathcal{W}}^{(lw)}[l]\right\} =(N-l)\mathcal{W}[l]\sigma^{2}(\delta_{0l}-1/N)\end{equation}
so the bias is

\myMathBox{\begin{equation}
\mathrm{Bias}\left\{ \widehat{C}_{\epsilon;\mathcal{W}}^{(lw)}[l]\right\} =\sigma^{2}\left((N\mathcal{W}[0]-1)\delta_{0,l}-\frac{(N-l)\mathcal{W}[l]}{N}\right)\end{equation}

}

\subsubsection{Covariance}

We start by deriving an expression for the covariance of the correlogram
for an arbitrary signal $Z$\begin{align*}
 & \frac{1}{\mathcal{W}[l]\mathcal{W}[l']}\mathrm{Cov}\left\{ \widehat{C}_{Z}^{(lw)}[l],\widehat{C}_{Z}^{(lw)*}[l']\right\} =\\
= & {\displaystyle \sum_{k=1}^{N-l}}{\displaystyle \sum_{k'=1}^{N-l'}}\mathrm{Cov}\left\{ Z^{'}[k]Z^{'*}[k+l]-\overline{Z^{'}}^{*}Z^{'}[k]-\overline{Z^{'}}Z^{'*}[k+l]+|\overline{Z^{'}}|^{2},\right.\\
 & \left.Z^{'*}[k']Z^{'}[k'+l']-\overline{Z^{'}}Z^{'*}[k']-\overline{Z^{'}}^{*}Z^{'}[k'+l']+|\overline{Z^{'}}|^{2}\right\} =\\
= & {\displaystyle \sum_{k=1}^{N-l}}{\displaystyle \sum_{k'=1}^{N-l'}}\left(\mathrm{Cov}\left\{ Z^{'}[k]Z^{'*}[k+l],Z^{'*}[k']Z^{'}[k'+l']\right\} -\mathrm{Cov}\left\{ Z^{'}[k]Z^{'*}[k+l],\overline{Z^{'}}Z^{'*}[k']\right\} +\right.\\
 & -\mathrm{Cov}\left\{ Z^{'}[k]Z^{'*}[k+l],\overline{Z^{'}}^{*}Z^{'}[k'+l']\right\} +\mathrm{Cov}\left\{ Z^{'}[k]Z^{'*}[k+l],|\overline{Z^{'}}|^{2}\right\} -\mathrm{Cov}\left\{ \overline{Z^{'}}^{*}Z^{'}[k],Z^{'*}[k']Z^{'}[k'+l']\right\} +\\
 & +\mathrm{Cov}\left\{ \overline{Z^{'}}^{*}Z^{'}[k],\overline{Z^{'}}Z^{'*}[k']\right\} +\mathrm{Cov}\left\{ \overline{Z^{'}}^{*}Z^{'}[k],\overline{Z^{'}}^{*}Z^{'}[k'+l']\right\} -\mathrm{Cov}\left\{ \overline{Z^{'}}^{*}Z^{'}[k],|\overline{Z^{'}}|^{2}\right\} +\\
 & -\mathrm{Cov}\left\{ \overline{Z^{'}}Z^{'*}[k+l],Z^{'*}[k']Z^{'}[k'+l']\right\} +\mathrm{Cov}\left\{ \overline{Z^{'}}Z^{'*}[k+l],\overline{Z^{'}}Z^{'*}[k']\right\} +\mathrm{Cov}\left\{ \overline{Z^{'}}Z^{'*}[k+l],\overline{Z^{'}}^{*}Z^{'}[k'+l']\right\} +\\
 & -\mathrm{Cov}\left\{ \overline{Z^{'}}Z^{'*}[k+l],|\overline{Z^{'}}|^{2}\right\} +\mathrm{Cov}\left\{ |\overline{Z^{'}}|^{2},Z^{'*}[k']Z^{'}[k'+l']\right\} -\mathrm{Cov}\left\{ |\overline{Z^{'}}|^{2},\overline{Z^{'}}Z^{'*}[k']\right\} +\\
 & \left.-\mathrm{Cov}\left\{ |\overline{Z^{'}}|^{2},\overline{Z^{'}}^{*}Z^{'}[k'+l']\right\} +\mathrm{Cov}\left\{ |\overline{Z^{'}}|^{2},|\overline{Z^{'}}|^{2}\right\} \right)\end{align*}
where $Z':=Z-\mu$.

The last expression contains 16 terms involving various covariances
of a general signal $Z$. To progress further we specialise $Z$ to
be the higher-order white noise signal $\epsilon$. Working out each
of the 16 covariance terms individually in the order the appear above
for this case we have: first covariance term\begin{align*}
\mathrm{Cov}\left\{ \epsilon^{'}[k]\epsilon^{'*}[k+l],\epsilon^{'*}[k']\epsilon^{'}[k'+l']\right\}  & =\delta_{0l}\delta_{0l'}\delta_{kk'}(\mu_{4}-\sigma^{4})+(1-\delta_{0l})\delta_{kk'}\delta_{ll'}\sigma^{4}+(1-\delta_{0l}\delta_{0l'})\delta_{k,k'+l'}\delta_{k',k+l}|m_{2}|^{2}=\\
 & =\delta_{0l}\delta_{0l'}\delta_{kk'}(\mu_{4}-2\sigma^{4}-|m_{2}|^{2})+\delta_{kk'}\delta_{ll'}\sigma^{4}+\delta_{k,k'+l'}\delta_{k',k+l}|m_{2}|^{2}=\\
 & =\delta_{0l}\delta_{0l'}\delta_{kk'}\kappa_{4}+\delta_{kk'}\delta_{ll'}\sigma^{4}+\delta_{k',k+l}\delta_{-l,l'}|m_{2}|^{2}\end{align*}
second covariance term\begin{align*}
\mathrm{Cov}\left\{ \epsilon^{'}[k]\epsilon^{'*}[k+l],\overline{\epsilon^{'}}\epsilon^{'*}[k']\right\}  & =\frac{1}{N}(\delta_{0l}\delta_{kk'}(\mu_{4}-\sigma^{4})+(1-\delta_{0l})(\delta_{kk'}\sigma^{4}+\delta_{k',k+l}|m_{2}|^{2}))=\\
 & =\frac{1}{N}(\delta_{0l}\delta_{kk'}(\mu_{4}-2\sigma^{4}-|m_{2}|^{2})+\delta_{kk'}\sigma^{4}+\delta_{l,k'-k}|m_{2}|^{2})=\\
 & =\frac{1}{N}(\delta_{0l}\delta_{kk'}\kappa_{4}+\delta_{kk'}\sigma^{4}+\delta_{k',k+l}|m_{2}|^{2})\end{align*}
 third covariance term\begin{align*}
\mathrm{Cov}\left\{ \epsilon^{'}[k]\epsilon^{'*}[k+l],\overline{\epsilon^{'}}^{*}\epsilon^{'}[k'+l']\right\}  & =\frac{1}{N}(\delta_{0l}\delta_{k,k'+l'}(\mu_{4}-\sigma^{4})+(1-\delta_{0l})(\delta_{k+l,k'+l'}\sigma^{4}+\delta_{k,k'+l'}|m_{2}|^{2}))=\\
 & =\frac{1}{N}(\delta_{0l}\delta_{k,k'+l'}(\mu_{4}-2\sigma^{4}-|m_{2}|^{2})+\delta_{k+l,k'+l'}\sigma^{4}+\delta_{k,k'+l'}|m_{2}|^{2})=\\
 & =\frac{1}{N}(\delta_{0l}\delta_{k,k'+l'}\kappa_{4}+\delta_{k+l,k'+l'}\sigma^{4}+\delta_{k,k'+l'}|m_{2}|^{2})\end{align*}
fourth covariance term\begin{align*}
\mathrm{Cov}\left\{ \epsilon^{'}[k]\epsilon^{'*}[k+l],|\overline{\epsilon^{'}}|^{2}\right\}  & =\frac{1}{N^{2}}(\delta_{0l}(\mu_{4}-\sigma^{4})+(1-\delta_{0l})(\sigma^{4}+|m_{2}|^{2}))=\\
 & =\frac{1}{N^{2}}(\delta_{0l}(\mu_{4}-2\sigma^{4}-|m_{2}|^{2})+\sigma^{4}+|m_{2}|^{2})=\\
 & =\frac{1}{N^{2}}(\delta_{0l}\kappa_{4}+\sigma^{4}+|m_{2}|^{2})\end{align*}
fifth covariance term\begin{align*}
\mathrm{Cov}\left\{ \overline{\epsilon^{'}}^{*}\epsilon^{'}[k],\epsilon^{'*}[k']\epsilon^{'}[k'+l']\right\}  & =\frac{1}{N}(\delta_{0l'}\delta_{kk'}(\mu_{4}-\sigma^{4})+(1-\delta_{0l'})(\delta_{kk'}\sigma^{4}+\delta_{k,k'+l'}|m_{2}|^{2}))=\\
 & =\frac{1}{N}(\delta_{0l'}\delta_{kk'}(\mu_{4}-2\sigma^{4}-|m_{2}|^{2})+(\delta_{kk'}\sigma^{4}+\delta_{k,k'+l'}|m_{2}|^{2}))=\\
 & =\frac{1}{N}(\delta_{0l'}\delta_{kk'}\kappa_{4}+\delta_{kk'}\sigma^{4}+\delta_{k,k'+l'}|m_{2}|^{2})\end{align*}
sixth covariance term\begin{align*}
\mathrm{Cov}\left\{ \overline{\epsilon^{'}}^{*}\epsilon^{'}[k],\overline{\epsilon^{'}}\epsilon^{'*}[k']\right\}  & =\frac{1}{N^{2}}(\delta_{kk'}(\mu_{4}-\sigma^{4}+(N-1)\sigma^{4})+(1-\delta_{kk'})|m_{2}|^{2})=\\
 & =\frac{1}{N^{2}}(\delta_{kk'}(\kappa_{4}+N\sigma^{4})+|m_{2}|^{2})\end{align*}
seventh\begin{align*}
\mathrm{Cov}\left\{ \overline{\epsilon^{'}}^{*}\epsilon^{'}[k],\overline{\epsilon^{'}}^{*}\epsilon^{'}[k'+l']\right\}  & =\frac{1}{N^{2}}(\delta_{k,k'+l'}(\mu_{4}-\sigma^{4}+(N-1)|m_{2}|^{2})+(1-\delta_{k,k'+l'})\sigma^{4})=\\
 & =\frac{1}{N^{2}}(\delta_{k,k'+l'}(\mu_{4}-2\sigma^{4}+(N-1)|m_{2}|^{2})+\sigma^{4})=\\
 & =\frac{1}{N^{2}}(\delta_{k,k'+l'}(\kappa_{4}+N|m_{2}|^{2})+\sigma^{4})\end{align*}
eighth\begin{align*}
\mathrm{Cov}\left\{ \overline{\epsilon^{'}}^{*}\epsilon^{'}[k],|\overline{\epsilon^{'}}|^{2}\right\}  & =\frac{1}{N^{3}}((\mu_{4}-\sigma^{4})+(N-1)(\sigma^{4}+|m_{2}|^{2}))=\\
 & =\frac{1}{N^{3}}((\mu_{4}-2\sigma^{4}-|m_{2}|^{2})+N(\sigma^{4}+|m_{2}|^{2}))=\\
 & =\frac{\kappa_{4}}{N^{3}}+\frac{\sigma^{4}+|m_{2}|^{2}}{N^{2}}\end{align*}
ninth\begin{align*}
\mathrm{Cov}\left\{ \overline{\epsilon^{'}}\epsilon^{'*}[k+l],\epsilon^{'*}[k']\epsilon^{'}[k'+l']\right\}  & =\frac{1}{N}(\delta_{k',k+l}\delta_{0,l'}(\mu_{4}-\sigma^{4})+(1-\delta_{0l'}\delta_{k',k+l})(\delta_{k+l,k'+l'}\sigma^{4}+\delta_{k',k+l}|m_{2}|^{2}))=\\
 & =\frac{1}{N}(\delta_{k',k+l}\delta_{0,l'}(\mu_{4}-2\sigma^{4}-|m_{2}|^{2})+\delta_{k+l,k'+l'}\sigma^{4}+\delta_{k',k+l}|m_{2}|^{2})=\\
 & =\frac{1}{N}(\delta_{k',k+l}\delta_{0,l'}\kappa_{4}+\delta_{k+l,k'+l'}\sigma^{4}+\delta_{k',k+l}|m_{2}|^{2})\end{align*}
tenth\begin{align*}
\mathrm{Cov}\left\{ \overline{\epsilon^{'}}\epsilon^{'*}[k+l],\overline{\epsilon^{'}}\epsilon^{'*}[k']\right\}  & =\frac{1}{N^{2}}(\delta_{k',k+l}(\mu_{4}-\sigma^{4}+(N-1)s^{4})+(1-\delta_{k',k+l})\sigma^{4})=\\
 & =\frac{1}{N^{2}}(\delta_{k',k+l}(\mu_{4}-2\sigma^{4}+(N-1)s^{4})+\sigma^{4})=\\
 & =\frac{1}{N^{2}}(\delta_{k',k+l}(\kappa_{4}+N|m_{2}|^{2})+\sigma^{4})\end{align*}
11-th\begin{align*}
\mathrm{Cov}\left\{ \overline{\epsilon^{'}}^{*}\epsilon^{'}[k],\overline{\epsilon^{'}}\epsilon^{'*}[k']\right\}  & =\frac{1}{N^{2}}(\delta_{kk'}(\mu_{4}-\sigma^{4}+(N-1)\sigma^{4})+(1-\delta_{kk'})|m_{2}|^{2})=\\
 & =\frac{1}{N^{2}}(\delta_{kk'}(\mu_{4}-\sigma^{4}-|m_{2}|^{2}+(N-1)\sigma^{4})+|m_{2}|^{2})=\\
 & =\frac{1}{N^{2}}(\delta_{kk'}(\kappa_{4}+N\sigma^{4})+|m_{2}|^{2})\end{align*}
12-th\begin{align*}
\mathrm{Cov}\left\{ \overline{\epsilon^{'}}\epsilon^{'*}[k+l],|\overline{\epsilon^{'}}|^{2}\right\}  & =\frac{1}{N^{3}}((\mu_{4}-\sigma^{4})+(N-1)(\sigma^{4}+|m_{2}|^{2}))=\\
 & =\frac{1}{N^{3}}((\mu_{4}-2\sigma^{4}-|m_{2}|^{2})+N(\sigma^{4}+|m_{2}|^{2}))=\\
 & =\frac{\kappa_{4}}{N^{3}}+\frac{\sigma^{4}+|m_{2}|^{2}}{N^{2}}\end{align*}
13-th\begin{align*}
\mathrm{Cov}\left\{ |\overline{\epsilon^{'}}|^{2},\epsilon^{'*}[k']\epsilon^{'}[k'+l']\right\}  & =\frac{1}{N^{2}}(\delta_{0l'}(\mu_{4}-\sigma^{4})+(1-\delta_{0l'})(\sigma^{2}+|m_{2}|^{2}))=\\
 & =\frac{1}{N^{2}}(\delta_{0l'}(\mu_{4}-2\sigma^{4}-|m_{2}|^{2})+(\sigma^{2}+|m_{2}|^{2}))=\\
 & =\frac{1}{N^{2}}(\delta_{0l'}\kappa_{4}+\sigma^{2}+|m_{2}|^{2})\end{align*}
14-th\begin{align*}
\mathrm{Cov}\left\{ |\overline{\epsilon^{'}}|^{2},\overline{\epsilon^{'}}\epsilon^{'*}[k']\right\}  & =\frac{1}{N^{3}}((\mu_{4}-\sigma^{4})+(N-1)(\sigma^{4}+|m_{2}|^{2}))=\\
 & =\frac{1}{N^{3}}((\mu_{4}-2\sigma^{4}-|m_{2}|^{2})+N(\sigma^{4}+|m_{2}|^{2}))=\\
 & =\frac{\kappa_{4}}{N^{3}}+\frac{\sigma^{4}+|m_{2}|^{2}}{N^{2}}\end{align*}
15-th\begin{align*}
\mathrm{Cov}\left\{ |\overline{\epsilon^{'}}|^{2},\overline{\epsilon^{'}}^{*}\epsilon^{'}[k'+l']\right\}  & =\frac{1}{N^{3}}((\mu_{4}-\sigma^{4})+(N-1)(\sigma^{4}+|m_{2}|^{2}))=\\
 & =\frac{1}{N^{3}}((\mu_{4}-2\sigma^{4}-|m_{2}|^{2})+N(\sigma^{4}+|m_{2}|^{2}))=\\
 & =\frac{\kappa_{4}}{N^{3}}+\frac{\sigma^{4}+|m_{2}|^{2}}{N^{2}}\end{align*}
and finally the last term\begin{align*}
\mathrm{Cov}\left\{ |\overline{\epsilon^{'}}|^{2},|\overline{\epsilon^{'}}|^{2}\right\}  & =E\left\{ |\overline{\epsilon^{'}}|^{4}\right\} -E\left\{ |\overline{\epsilon^{'}}|^{2}\right\} E\left\{ |\overline{\epsilon^{'}}|^{2}\right\} =\\
 & =\frac{1}{N^{4}}(N\mu_{4}+N(N-1)(2\sigma^{4}+|m_{2}|^{2}))-\frac{\sigma^{4}}{N^{2}}=\\
 & =\frac{1}{N^{3}}(\mu_{4}+(N-1)(2\sigma^{4}+|m_{2}|^{2})-N\sigma^{4})=\\
 & =\frac{\kappa_{4}}{N^{3}}+\frac{\sigma^{4}+|m_{2}|^{2}}{N^{2}}\end{align*}
for $l\ge0,\; l'\ge0$. Adding up all the terms we can now collect
all terms containing the lag dependent delta functions$[\delta_{0l}\delta_{0l'},(\delta_{0l}+\delta_{0l'}),\delta_{l,l'},1]$.
The term containing $\delta_{0l}\delta_{0l'}$ is\[
+\delta_{0l}\delta_{0l'}(\kappa_{4}+s^{4})\delta_{k,k'}\]
the terms containing $\delta_{0l}$ and $\delta_{0l'}$ are\[
-\frac{\kappa_{4}}{N}\left((\delta_{k,k'}+\delta_{k,k'+l'}-\frac{1}{N})\delta_{0l}+(\delta_{k,k'}+\delta_{k',k+l}-\frac{1}{N})\delta_{0l'}\right)\]
the terms containing $\delta_{l,l'}$ is\[
+\delta_{l,l'}(\sigma^{4}\delta_{k,k'})\]
and all the other terms are\begin{align*}
 & -\frac{1}{N}\left(2\delta_{k+l,k'+l'}\sigma^{4}+(\delta_{k,k'+l'}+\delta_{k',k+l})s^{4}+2\delta_{k,k'}\sigma^{4}+(\delta_{k,k'+l'}+\delta_{k',k+l})s^{4}-2\frac{\kappa_{4}+N\sigma^{4}}{N}\delta_{k,k'}+\right.\\
 & \left.\quad-(\delta_{k,k'+l'}+\delta_{k',k+l})\frac{\kappa_{4}+Ns^{4}}{N}-2\frac{\sigma^{4}+s^{4}}{N}-2\frac{\sigma^{4}+s^{4}}{N}+3\frac{\sigma^{4}+s^{4}}{N}+3\frac{\kappa_{4}}{N^{2}}\right)\\
= & -\frac{1}{N}\left((2\delta_{k,k'}+2\delta_{k+l,k'+l'}-2\delta_{kk'}+\frac{-2-2+3}{N})\sigma^{4}+\right.\\
 & \left.\quad+(2\delta_{k,k'+l'}+2\delta_{k',k+l}-\delta_{k,k'+l'}-\delta_{k',k+l}+\frac{-2-2+3}{N})s^{4}-\frac{\kappa_{4}}{N}(2\delta_{k,k'}+\delta_{k,k'+l'}+\delta_{k',k+l}-\frac{3}{N})\right)\\
= & -\frac{1}{N}\left((2\delta_{k+l,k'+l'}-\frac{1}{N})\sigma^{4}+(\delta_{k,k'+l'}+\delta_{k',k+l}-\frac{1}{N})s^{4}-\frac{\kappa_{4}}{N}(2\delta_{k,k'}+\delta_{k,k'+l'}+\delta_{k',k+l}-\frac{3}{N})\right)\end{align*}
Now we perform the double sum over all $k,k'$ each of the lag dependent
delta terms individually and make use of the summation formulas (\ref{eq:LWsum_diag}),
(\ref{eq:LWsum_offdiag}), (\ref{eq:LWsum_offdiagprim}), (\ref{eq:LWsum_2offdiag}),
and (\ref{eq:LWsum_2xUnity}). First we have a contribution to the
zero lag variance, $\delta_{0l}\delta_{0l'}$, which becomes\[
+\delta_{0l}\delta_{0l'}(\kappa_{4}+s^{4})(N-\max(l,l'))\]
then a contribution to the zero lag versus any lag covariance terms,
$\delta_{0l}$ and $\delta_{0l'}$, which become\begin{multline*}
-\frac{\kappa_{4}}{N}\left((2N-\max(l,l')-l-l'-\frac{(N-l)(N-l')}{N})\delta_{0l}+(2N-\max(l,l')-l-l'-\frac{(N-l)(N-l')}{N})\delta_{0l'}\right)=\\
=-\frac{\kappa_{4}}{N}\left((N-l')\delta_{0l}+(N-l)\delta_{0l'}\right)=-\kappa_{4}\left(\left(1-\frac{l'}{N}\right)\delta_{0l}+\left(1-\frac{l}{N}\right)\delta_{0l'}\right)\end{multline*}
then a contribution to the any lag variance term, $\delta_{l,l'}$,
which becomes\[
+\delta_{l,l'}\sigma^{4}(N-\max(l,l'))\]
while all other terms become\begin{align*}
 & -\frac{1}{N}\left((2(N-\max(l,l'))-\frac{(N-l)(N-l')}{N})\sigma^{4}+2(N-\min(l+l',N))-\frac{(N-l)(N-l')}{N})s^{4}\right.+\\
 & \quad\left.-\frac{\kappa_{4}}{N}(2(N-\max(l,l'))+2(N-\min(l+l',N))-3\frac{(N-l)(N-l')}{N})\right)=\\
= & -\frac{1}{N}\left((N-2\max(l,l')+l+l'-\frac{ll'}{N})\sigma^{4}+(N-2\min(l+l',N))+l+l'-\frac{ll'}{N})s^{4}\right.+\\
 & \quad\left.-\frac{\kappa_{4}}{N}(N-2\max(l,l')-2\min(l+l',N)+3(l+l'-\frac{ll'}{N}))\right)\end{align*}

Finally, adding up the individual terms, we arrive at the final expression
for the covariance. For $ll'\ge0$, it is

\myMathBox{\begin{multline}
\frac{1}{\mathcal{W}[l]\mathcal{W}[l']}\mathrm{Cov}\left\{ \widehat{C}_{\epsilon}^{(lw)}[l],\widehat{C}_{\epsilon}^{(lw)}[l']\right\} =\delta_{0,l}\delta_{0,l'}(\kappa_{4}+s^{4})N-\delta_{0,l}\kappa_{4}\left(1-\frac{|l'|}{N}\right)-\delta_{0,l'}\kappa_{4}\left(1-\frac{|l|}{N}\right)\\
+\delta_{l,l'}\sigma^{4}(N-|l|)+\kappa_{4}\left(\frac{1}{N}-\frac{2\max(|l|,|l'|)+2\min(|l|+|l'|,N)-3(|l|+|l'|)}{N^{2}}-3\frac{ll'}{N^{3}}\right)+\\
-\sigma^{4}\left(1-\frac{2\max(|l|,|l'|)-|l|-|l'|}{N}-\frac{ll'}{N^{2}}\right)-s^{4}\left(1-\frac{2\min(|l|+|l'|,N)-|l|-|l'|}{N}-\frac{ll'}{N^{2}}\right),\\
ll'\ge0.\end{multline}
} These results are comparable to \cite{Anderson71} Theorem 8.2.6
for the case of real white noise.

For $ll'\le0$, the analogous expression for the covariance is\begin{multline}
\frac{1}{\mathcal{W}[l]\mathcal{W}[l']}\mathrm{Cov}\left\{ \widehat{C}_{\epsilon}^{(lw)}[l],\widehat{C}_{\epsilon}^{(lw)}[l']\right\} =\delta_{0,l}\delta_{0,l'}(\kappa_{4}+s^{4})N-\delta_{0,l}\kappa_{4}\left(1-\frac{|l'|}{N}\right)-\delta_{0,l'}\kappa_{4}\left(1-\frac{|l|}{N}\right)\\
+\delta_{|l||l'|}s^{4}(N-|l|)+\kappa_{4}\left(\frac{1}{N}-\frac{2\max(|l|,|l'|)+2\min(|l|+|l'|,N)-3(|l|+|l'|)}{N^{2}}-3\frac{ll'}{N^{3}}\right)+\\
-\sigma^{4}\left(1-\frac{2\max(|l|,|l'|)-|l|-|l'|}{N}-\frac{ll'}{N^{2}}\right)-s^{4}\left(1-\frac{2\min(|l|+|l'|,N)-|l|-|l'|}{N}-\frac{ll'}{N^{2}}\right),\\
ll'\le0.\end{multline}

\subsubsection{Mean squared error}

From the bias and the covariances one can determine the mean square
error. From the definition of the MSE we find that\begin{align*}
 & \mathrm{MSE}\left\{ \widehat{C}_{\epsilon}^{(lw)}[l]\right\} =\mathrm{Var}\left\{ \widehat{C}_{\epsilon}^{(lw)}[l]\right\} +\left|\mathrm{Bias}\left\{ \widehat{C}_{\epsilon}^{(lw)}[l]\right\} \right|^{2}=\\
= & \mathcal{W}^{2}[l]\left(\delta_{0,l}\left((\kappa_{4}+s^{4})N-2\kappa_{4}\right)+\sigma^{4}(N-l)+\kappa_{4}\left(\frac{1}{N}-\frac{2l+2\min(2l,N)-6l}{N^{2}}-3\frac{l^{2}}{N^{3}}\right)+\right.\\
 & \left.-\sigma^{4}\left(1-\frac{l^{2}}{N^{2}}\right)-s^{4}\left(1-\frac{2\min(2l,N)-2l}{N}-\frac{l^{2}}{N^{2}}\right)\right)+\sigma^{4}\left|(N\mathcal{W}[0]-1)\delta_{0,l}-\frac{(N-l)\mathcal{W}[l]}{N}\right|^{2}=\\
= & \mathcal{W}^{2}[l]\left(\delta_{0,l}((\kappa_{4}+s^{4})N-2\kappa_{4})+\sigma^{4}(N-l-1+\frac{l^{2}}{N^{2}})+\kappa_{4}\left(\frac{1}{N}-\frac{2l+2\min(2l,N)-6l}{N^{2}}-3\frac{l^{2}}{N^{3}}\right)+\right.\\
 & \left.-s^{4}\left(1-\frac{2\min(2l,N)-2l}{N}-\frac{l^{2}}{N^{2}}\right)\right)+\sigma^{4}\left(\left((N-2)N\mathcal{W}^{2}[0]-2(N-1)\mathcal{W}[0]+1\right)\delta_{0,l}+\frac{(N-l)^{2}\mathcal{W}[l]^{2}}{N^{2}}\right)=\\
= & \delta_{0l}\left(\left((\kappa_{4}+s^{4})N-2\kappa_{4}\left(1-\frac{l}{N}\right)+\sigma^{4}(N-2)N\right)\mathcal{W}^{2}[l]-2(N-1)\sigma^{4}\mathcal{W}[0]+\sigma^{4}\right)+\\
 & +\left(\sigma^{4}\left(N-l-1+\frac{l^{2}}{N^{2}}\right)+\kappa_{4}\left(\frac{1}{N}-\frac{2l+2\min(2l,N)-6l}{N^{2}}-3\frac{l^{2}}{N^{3}}\right)+\right.\\
 & \left.-s^{4}\left(1-\frac{2\min(2l,N)-2l}{N}-\frac{l^{2}}{N^{2}}\right)+\frac{(N-l)^{2}\sigma^{4}}{N^{2}}\right)\mathcal{W}^{2}[l]\end{align*}
thus

\myMathBox{\begin{align}
\mathrm{MSE}\left\{ \widehat{C}_{\epsilon}^{(lw)}[0]\right\}  & =(\sigma^{4}(N-1)N+(\kappa_{4}+s^{4})N-s^{4}-2\kappa_{4}+\frac{\kappa_{4}}{N})\mathcal{W}^{2}[0]-2(N-1)\sigma^{4}\mathcal{W}[0]+\sigma^{4}\\
\mathrm{MSE}\left\{ \widehat{C}_{\epsilon}^{(lw)}[l\neq0]\right\}  & =\left(\sigma^{4}(N-l)(1-\frac{2l}{N^{2}})+\frac{\kappa_{4}}{N}\left(1-\frac{2l+2\min(2l,N)-6l}{N}-3\frac{l^{2}}{N^{2}}\right)\right.\nonumber \\
 & \quad\left.-s^{4}\left(1-\frac{2\min(2l,N)-2l}{N}-\frac{l^{2}}{N^{2}}\right)\right)\mathcal{W}^{2}[l]\end{align}
}

Asymptotically i.e. $N\rightarrow\infty$, keeping only terms of order
$N$ or $l$we find\begin{align}
\mathrm{MSE}\left\{ \widehat{C}_{\epsilon}^{(lw)}[0]\right\}  & =\sigma^{4}N^{2}\mathcal{W}^{2}[0]-2N\sigma^{4}\mathcal{W}[0]+\sigma^{4}\\
\mathrm{MSE}\left\{ \widehat{C}_{\epsilon}^{(lw)}[l\neq0]\right\}  & =\sigma^{4}(N-l)\mathcal{W}^{2}[l]\end{align}

\section{Conclusion}

We have derived the sampling properties up to second-order of the
ACS and ACVS estimators $\widehat{R}^{(lw)}$, $\widehat{R}^{(fl)}$
and $\widehat{C}^{(lw)}$ for a general white noise sequence $\epsilon[\cdot]$.
An interesting result we have found is that the covariances of the
correlograms in general have a lag dependence. This is despite the
fact that the noise sequence,$\epsilon[\cdot]$ for which these covariances
were derived, is not itself lag dependent.

Further conclusions based on the results derived here will be given
in a following paper \cite{Carozzi05b}.

\appendix

\section{Summation formulas}

In deriving the second order sampling properties we have used the
following summation formulas for the lag windowed correlograms (with
and without mean removal) in sections \ref{sub:Lag-win-ACF-est} and
\ref{sub:ACVF-est},

\begin{align}
{\displaystyle \sum_{k=1}^{N-|l|}}{\displaystyle \sum_{k'=1}^{N-|l'|}}\delta_{k,k'} & =N-\max(|l|,|l'|)\quad0\le|l|\le N,\,0\le|l'|\le N\label{eq:LWsum_diag}\\
{\displaystyle \sum_{k=1}^{N-|l|}}{\displaystyle \sum_{k'=1}^{N-|l'|}}\delta_{k',k+|l|} & =N-\min(|l|+|l'|,N)\quad0\le|l|\le N,\,0\le|l'|\le N\label{eq:LWsum_offdiag}\\
{\displaystyle \sum_{k=1}^{N-|l|}}{\displaystyle \sum_{k'=1}^{N-|l'|}}\delta_{k,k'+|l'|} & =N-\min(|l|+|l'|,N)\quad0\le|l|\le N,\,0\le|l'|\le N\label{eq:LWsum_offdiagprim}\\
{\displaystyle \sum_{k=1}^{N-|l|}}{\displaystyle \sum_{k'=1}^{N-|l'|}}\delta_{k+|l|,k'+|l'|} & =N-\max(|l|,|l'|)\quad0\le|l|\le N,\,0\le|l'|\le N\label{eq:LWsum_2offdiag}\\
{\displaystyle \sum_{k=1}^{N-|l|}}{\displaystyle \sum_{k'=1}^{N-|l'|}}1 & =(N-|l|)(N-|l'|)\quad0\le|l|\le N,\,0\le|l'|\le N\label{eq:LWsum_2xUnity}\end{align}

The following sums were used with the fixed-length summation estimator
in section \ref{sub:Fix-len-ACF-est},

\begin{align}
{\displaystyle \sum_{k=1}^{L}}{\displaystyle \sum_{k'=1}^{L}}\delta_{k,k'} & =L\quad0\le|l|\le M,\,0\le|l'|\le M\label{eq:FLsum_diag}\\
{\displaystyle \sum_{k=1}^{L}}{\displaystyle \sum_{k'=1}^{L}}\delta_{k',k+|l|} & =L-\min(|l|,L)\quad0\le|l|\le M,\,0\le|l'|\le M\label{eq:FLsum_offdiag}\\
{\displaystyle \sum_{k=1}^{L}}{\displaystyle \sum_{k'=1}^{L}}\delta_{k,k'+|l'|} & =L-\min(|l'|,L)\quad0\le|l|\le M,\,0\le|l'|\le M\label{eq:FLsum_offdiagprim}\\
{\displaystyle \sum_{k=1}^{L}}{\displaystyle \sum_{k'=1}^{L}}\delta_{k+|l|,k'+|l'|} & =L-\min\left(\left||l|-|l'|\right|,L\right)\quad0\le|l|\le M,\,0\le|l'|\le M\label{eq:FLsum_2offdiag}\\
{\displaystyle \sum_{k=1}^{L}}{\displaystyle \sum_{k'=1}^{L}}1 & =L^{2}\quad0\le|l|\le M,\,0\le|l'|\le M\label{eq:FLsum_2xUnity}\end{align}
where $M:=N-L$.

\section*{Acknowledgments}

This work was sponsored by PPARC ref: PPA/G/S/1999/00466 and PPA/G/S/2000/00058.\bibliographystyle{plain}
\bibliography{main}

\begin{thebibliography}{10}

\bibitem{Anderson71}
T.~W. Anderson.
\newblock {\em The Statistical Analysis of Time Series}.
\newblock John Wiley \& Sons, Inc., 1971.

\bibitem{Bartlett46}
M.~S. Bartlett.
\newblock On the theoretical specification and sampling properties of
  autocorrelated time-series.
\newblock {\em Supplement to the Journal of the Royal Statistical Society},
  8(1):27--41, 1946.

\bibitem{Bendat00}
Juilus~S. Bendat and Allan~G. Piersol.
\newblock {\em Random data: analysis and measurement procedures}.
\newblock John Wiley and Sons, Inc, third edition edition, 2000.

\bibitem{Bertorelle95}
Giorgio Bertorelle and Guido Barbujani.
\newblock Analysis of dna diversity by spatial autocorrelation.
\newblock {\em Genetics}, 140:811--819, 1995.

\bibitem{Blackman58}
R.~B. Blackman and J.~W. Tukey.
\newblock {\em The measurement of power spectra: from the point of view of
  communications engineering}.
\newblock Dover Publications, Inc, 1958.

\bibitem{Buckley2000}
A.~M. Buckley, M.~P. Gough, H.~Alleyne, K.~Yearby, and I.~Willis.
\newblock Measurement of wave-paricle interaction in the magnetosphere using
  the particle correlator experiments on cluster.
\newblock {\em European Space Agency}, ESA SP-449:303--306, 2000.

\bibitem{Carozzi05b}
T.~D. Carozzi and A.~M. Buckley.
\newblock Sampling errors of correlograms with and without sample mean removal
  for higher-order complex white noise with arbitrary mean.
\newblock {\em To be published in Journal of Time Series Analysis}, 2005.

\bibitem{Gough03}
M.~P. Gough, A.~M. Buckley, T.~Carozzi, and N.~Beloff.
\newblock Experimental studies of wave-particle interactions in space using
  particle correlators: results and future developments.
\newblock {\em Adv.\ Space Res.}, 32(3):407--416, 2003.

\bibitem{Marriott54}
F.~H.~C. Marriott and J.~A. Pope.
\newblock Bias in the estimation of autocorrelations.
\newblock {\em Biometrika}, 41(3/4):390--402, 1954.

\bibitem{Percival93a}
Donald~B. Percival and Andrew~T. Walden.
\newblock {\em Spectral Analysis for Physical Applications: multitaper and
  conventional univariate techniques}.
\newblock Cambridge University Press, 1993.

\bibitem{Schiffler96}
Andreas Schiffler.
\newblock {\em Superdarn measurements of double-peaked velocity spectra}.
\newblock PhD thesis, University of Saskatchewan, Canada, 1996.

\bibitem{Smolders99}
A.~B. Smolders and M.~P. van Haarlem, editors.
\newblock {\em Perspectives on Radio Astronomy: Technologies for Large Antenna
  Arrays}, Dwingeloo, The Netherlands, April 1999. ASTRON.

\bibitem{Weinreb63}
Sander Weinreb.
\newblock {\em A Digital Spectral Analysis Technique and Its Application to
  Radio Astronomy}.
\newblock PhD thesis, Massachusetts institute of technology, 1963.

\end{thebibliography}

\end{document}